\documentclass[lettersize,journal]{IEEEtran}
\usepackage{amsmath,amsfonts}
\usepackage{hyperref}
\usepackage{stfloats}
\usepackage{algorithm}
\usepackage{algorithmic}
\usepackage{array}
\usepackage{tikz}
\newcolumntype{C}[1]{>{\centering\let\newline\\\arraybackslash\hspace{0pt}}m{#1}}
\usepackage{graphicx}
\def\BibTeX{{\rm B\kern-.05em{\sc i\kern-.025em b}\kern-.08em
    T\kern-.1667em\lower.7ex\hbox{E}\kern-.125emX}}

\usetikzlibrary{positioning}
\usetikzlibrary{calc}
\usetikzlibrary{decorations,decorations.pathmorphing,decorations.text}
\usetikzlibrary{fit}
\usetikzlibrary{arrows.meta}
\usetikzlibrary{fadings, shadings}
\usetikzlibrary{patterns, patterns.meta}
\usetikzlibrary{shapes.geometric}

\title{Time Synchronization of TESLA-enabled GNSS Receivers}

\author{
    Jason~Anderson, Sherman~Lo, Todd~Walter \textit{Stanford~University}
    }

\begin{document}

\maketitle

\begin{abstract}
    As TESLA-enabled GNSS for authenticated positioning reaches ubiquity, receivers must use an onboard, GNSS-independent clock (GIC) and carefully constructed time synchronization algorithms to assert the authenticity afforded.
    This work provides the necessary checks and synchronization protocols needed in the broadcast-only GNSS context.
    We provide proof of security for each of our algorithms under a delay-capable adversary.
    The algorithms included herein enable a GNSS receiver to use its GIC to determine whether a message arrived at the correct time, to determine whether its GIC is safe to use and when the clock will no longer be safe in the future due to predicted clock drift, and to resynchronize its GIC.
    Each algorithm is safe to use even when an adversary induces delays within the protocol.
    Moreover, we discuss the implications of GNSS authentication schemes that use two simultaneous TESLA instances of different authentication cadences.
    To a receiver implementer or standards author, this work provides the necessary implementation algorithms to assert security and provides a comprehensive guide on why these methods are required.
    We discuss and address a vulnerability related to the standard synchronization protocols in the context of broadcast-only TESLA.
\end{abstract}


\section{Introduction}
    
    \IEEEPARstart{G}{lobal} Navigation Satellite System (GNSS) signals include two layered digital components: (1) the spreading code and (2) the navigation data.
    The spreading code layer provides the ranging signal, allowing users to deduce their range to a satellite.
    The navigation data layer provides users with satellite positions and other constellation data.
    From the ranges and satellite data, users deduce their position and time.
    This work pertains to secure receiver time synchronization to establish trust in the received GNSS navigation data authenticated with Timed Efficient Stream Loss-tolerant Authentication (TESLA).

    Several methods exist to augment spreading codes with authentication, including (1) watermarking the spreading code with cryptographic pseudorandom punctures (e.g., \cite{Anderson2017,anderson2022efficient}) and (2) encrypting the signal and employing a delay-release schedule of the encryption keys \cite{anderson2022cryptographic}.
    For the navigation data, one method is already in practice \cite{galIgnacio, Galicd, Gal2021}.
    And another is under consideration~\cite{AndersonnJournal}.
    Challenges hinder the quick augmentation of GNSS with authentication cryptography, such as bandwidth limitation and maintaining backward compatibility.
    Within all of the methods above, a receiver must have an onboard, GNSS-independent clock (GIC) that employs some non-GNSS, two-way (e.g., network) time synchronization.
    The GIC cannot be modified by the timing solution provided by GNSS, or any other broadcast-only signal.
    Otherwise, the receiver will not reject forged messages transmitted by a spoofer with a significant delay.
    Moreover, the GIC must be an always-on, real-time clock if the receiver is turned off in between synchronizations.

    The receipt time of a GNSS signal determines the receiver-deduced position and time.
    If a man-in-the-middle repeats the GNSS signal to a receiver, the receiver will deduce a time estimate that lags behind authentic GNSS time.
    For instance, a spoofer could delay the GNSS signal, slowly increasing the delay over time until the receiver would accept a spoofed message under TESLA.
    Here lies the catch-22 that necessitates non-GNSS timing assistance for GNSS authentication security.
    Because of the broadcast-only nature of GNSS and how simply repeating a GNSS signal could be accepted by the receiver as a lagging signal, synchronization requires a non-GNSS, two-way, and recurring connection~\cite{securetimetransfer}.

    The receiver must have a GIC to reject such delayed signals.
    If that receiver GIC lags too much at any time, the security TESLA provides breaks, and the receiver will accept arbitrary forgeries.
    This work discusses how to (1) use the required GIC to assert authentication security, (2) assert a healthy (i.e., non-lagging) receiver GIC and when the clock will no longer be safe in the future due to expected GIC drift, and (3) safely resynchronize the GIC.
    For each algorithm, we prove the conditions of its security under an adversarial model.
    Lastly, we discuss the synchronization context when using multiple TESLA instances of different timing requirements.
    
\begin{figure}
        \centering
        \includegraphics[width=\linewidth]{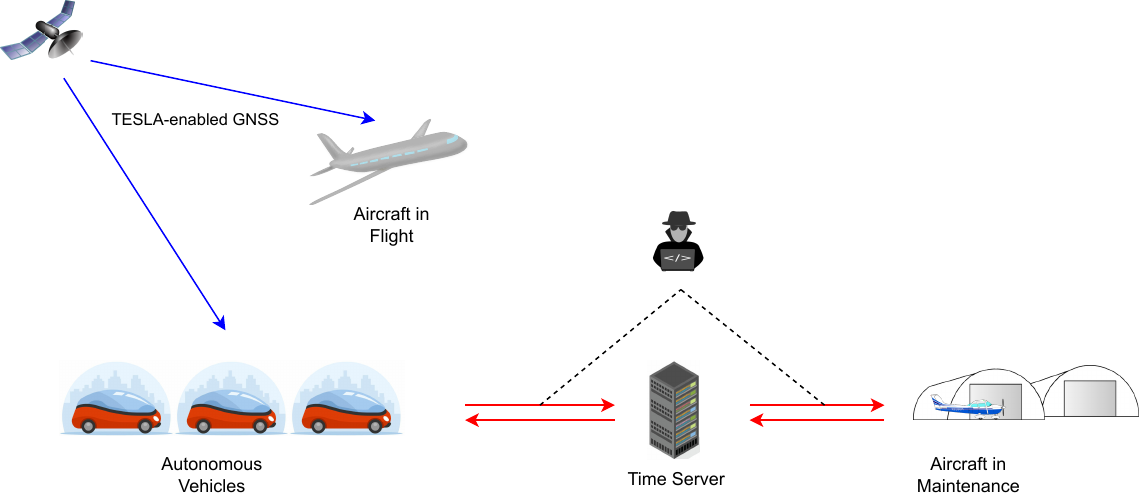}
        \caption{
            A conceptual diagram of our expectation of the future.
            We expect certain systems will need to use the authentication provided by current and future TESLA-enabled GNSS.
            Some receivers, such as autonomous cars, will have continuous access to a time server allowing for continuous checking to enforce the TESLA loose-time synchronization requirement.
            Other receivers, such as aircraft, might not have continuous access to a server but will use periodic maintenance and low-drift GICs between maintenance sessions to enforce the TESLA loose-time synchronization requirement.
            An adversary can delay the receipt times of synchronization signals to and from a synchronization server.
            }
        \label{fig: concept}
    \end{figure}

    To provide a concrete and plausible scenario and show the importance of this work, we discuss our vision and provide the conceptual diagram of Figure~\ref{fig: concept}.
    We envision a future where systems, particularly those connected with the safety of life, will want (or be required) to enforce the authentication of current and future GNSS signals.
    Users might have continuous access to an internet connection (e.g., autonomous cars) or not (e.g., aircraft).    
    Receivers may have access to GICs of varying drift rates.
    For instance, we expect receivers without an internet connection will have access to a low-drift GIC and need to conduct recurring clock checks during regular maintenance. 
    In contrast, receivers with an internet connection can frequently check their low-cost GIC.
    In the future, we expect millions of receivers will use the algorithms presented in this work to ensure the security of TESLA-enabled GNSS.

    \subsection{Notation}
        In this work, we will frequently describe the time of an event in two different clock frames: (1) the GNSS provider time and (2) the GIC of a particular GNSS receiver attempting to assert the authenticity of a TESLA-enabled GNSS signal.
        Let a particular synchronization be indexed by $l$, and let each synchronization include several sub-events indexed by $i$.
        An event $i$ during synchronization $l$ occurs at $t_i^l$ in the GNSS provider clock frame and $\tau_i^l$ in the receiver clock frame, via the measurement Equation~\eqref{eq: meas eqn} with $\theta^l$ as the receiver clock offset during synchronization $l$:
        \begin{equation} \label{eq: meas eqn}
            \tau_i^l = t_i^l + \theta^l.
        \end{equation}
        During $l$, we reasonably assume $\theta^l$ is constant.

        In Equation~\eqref{eq: meas eqn}, $\theta>0$ means the receiver clock runs {\em ahead} of the provider time.
        Authentication security will break when the receiver clock excessively {\em lags} provider time, and the procedures herein will provably determine or fix the health of a receiver clock provided there is no denial of service.

    \subsection{Network Time Security}

        Network Time Protocol (NTP) from \cite{rfcntp} provides a simple synchronization protocol between two clocks with a network connection.
        Immediately, we use its security extension, Network Time Security (NTS), which is NTP augmented with authentication cryptography~\cite{rfcnts}.
        We provide a brief algorithmic overview of NTS in Algorithm~\ref{alg: nts}.
        The methods herein can immediately extend to PTP (IEEE 1588-2019).
        We use NTS herein for simplicity and brevity.
        GNSS serves as an excellent proxy for reference time.
        NTP Stage 0 clocks are usually atomic clocks or GNSS (noting that GNSS is a collection of orbiting atomic clocks).
        UTC-GPS is tightly constrained to UTC-USNO, so GNSS is the reference time for this work.

        \begin{algorithm}
            \caption{Network Time Security (DO NOT USE FOR GNSS TESLA).}
            \label{alg: nts}
            \begin{algorithmic}[1]
                \STATE GNSS provider (or a delegate) and receiver establish an asymmetric authentication security instance.
                \STATE Receiver draws a nonce $\eta$.
                \STATE Receiver sends message $m_1 = (\eta, \tau_1, s^\textrm{receiver}_1)$ where $\tau_1$ is the time receiver recorded at the moment of sending $m_1$ and $s^\textrm{receiver}_1$ is receiver's authentication signature on $(\eta, \tau_1)$.
                \STATE Provider records $t_2$, the time of receipt of the message $m_1$.
                \STATE Provider sends message $m_2=(\eta, \tau_1, t_2, t_3, s^\textrm{provider}_2)$ back to receiver, where $s^\textrm{provider}_2$ is provider's authentication signature on $(\eta, \tau_1, t_2, t_3)$. $t_3$ is the moment that $m_2$ is transmitted back to the receiver.
                \STATE Receiver records $\tau_4$ at the moment of receipt of message $m_2$.
                \STATE The measured clock drift is $\hat \theta = \frac{1}{2}(\tau_1 -t_2 - t_3 + \tau_4)$ assuming that the transit time of $m_1$ and $m_2$ are the same.
            \end{algorithmic}
        \end{algorithm}

        \begin{figure}
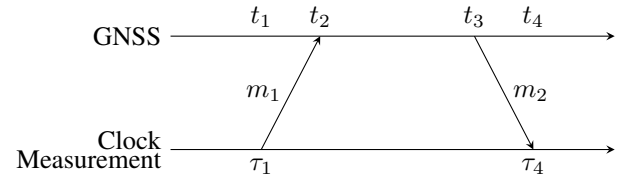

            \centering
            \tikzstyle{connect} = [inner sep=0pt, outer sep=0pt, circle, fill=black, minimum size=0pt]
            \def\lineLength{6}
            \def\eplength{0.6}
            \def\eventlength{1.2}   
            \tikz{
                \node[anchor=east] (GNSS) at (0, 0) {GNSS};
                \node[anchor=east, text width=2cm, align=right, font=\linespread{0.5}\selectfont] (RecC) at (0, -1.5) {Clock Measurement};

                \node (GNSSe) at (\lineLength, 0) {};
                \node (RecCe) at (\lineLength, -1.5) {};

                \draw[-stealth] (GNSS) -- (GNSSe);
                \draw[-stealth] (RecC) -- (RecCe);

                \node[connect] (t1) at (1*\eventlength, 0) {};
                \node[above] (lt1) at (t1) {$t_1$};

                \node[connect] (t2) at (2*\eventlength - 0.35*\eventlength, 0) {};
                \node[above] (lt2) at (t2) {$t_2$};

                \node[connect] (t3) at (3*\eventlength + 0.35*\eventlength, 0) {};
                \node[above] (lt3) at (t3) {$t_3$};

                \node[connect] (t4) at (4*\eventlength, 0) {};
                \node[above] (lt4) at (t4) {$t_4$};

                \node[connect] (tau1) at (1*\eventlength, -1.5) {};
                \node[below] (ltau1) at (tau1) {$\tau_1$};

                \node[connect] (tau4) at (4*\eventlength, -1.5) {};
                \node[below] (ltau4) at (tau4) {$\tau_4$};

                \draw[-stealth] (tau1) -- node[midway, left] {$m_1$} (t2);
                \draw[-stealth] (t3) -- node[midway, right] {$m_2$} (tau4);
            }
            \caption{A conceptual diagram of Algorithm~\ref{alg: nts}: NTS. The diagram depicts increasing time from left to right in the provider and receiver clock frames, and the messages shared between them. The protocol presumes provider and receiver have already established a cryptographic authentication instance to protect the {\em content} of the messages transmitted. The protocol is susceptible to adversary-induced delays in message transmission (like in Figure~\ref{fig: cert attack}.}
            \label{fig: nts}
        \end{figure}

        NTS provides authentication security on the {\em content} of the messages~\cite{rfcnts}.
        However, the {\em measured} sending and arrival times determine the estimated clock drift, and the adversary can manipulate the measured arrival times via man-in-the-middle delays.
        No cryptographic protection is available for message delays~\cite{futile}.
        There are several suggested mitigations to man-in-the-middle delays.
        These include querying multiple random NTS servers and rejecting synchronization with large NTS round-trip times.
        While helpful, they do not provide provable safety beyond the round-trip time uncertainty, and we are primarily concerned about {\em lagging} clocks in this work.

        NTS uses two messages: (1) a message from the receiver to the provider and (2) a message from the provider to the receiver.
        The times of sending and receipt of the two messages are referred to as synchronization events 1, 2, 3, and 4, respectively and determine the estimated receiver clock offset $\hat \theta$.
        These events are diagrammed in Figure~\ref{fig: nts}.
        The estimated $\hat \theta$ derives from the assumption that the transit times from receiver to provider and receiver to provider are the same, as in Equation~\eqref{eq: same transit time} to Equation~\eqref{eq: nts} via the measurement Equation~\eqref{eq: meas eqn}:
        \begin{align}
            t_2^l - t_1^l &= t_4^l - t_3^l, \label{eq: same transit time} \\
            t_2^l - (\tau_1^l - \hat \theta^l) &= (\tau_4^l - \hat \theta^l) - t_3^l, \nonumber \\
            \hat \theta^l &= \frac{1}{2}(\tau_1^l -t_2^l - t_3^l + \tau_4^l), \nonumber \\
            \hat \theta^l &= \frac{1}{2}(\tau_1^l + \tau_4^l) - \frac12(t_2^l + t_3^l). \label{eq: nts}
        \end{align}
        Equation~\eqref{eq: nts} is arranged in the form that corresponds to estimating the horizontal midpoint of the trapezoid from Figure~\ref{fig: nts}.

        While an adversary can delay the signals to produce an arbitrary $\hat\theta$ on the receiver, we can still provably bound $\hat\theta$ so as to notify the receiver of an insecure GIC.
        Under the reasonable assumption that an adversary cannot tamper directly with the relevant onboard GIC oscillators, we can assume that the measurement of each oscillator linearly increases, as in Equation~\eqref{eq: tau4>tau3} to derive Equation~\eqref{eq: theta upper bound} via measurement Equation~\eqref{eq: meas eqn}:
        \begin{align}
            0 &< \tau_4^l - \tau_3^l \label{eq: tau4>tau3}, \\
            0 &< \tau_4^l - (t_3^l + \theta^l),  \nonumber \\
            \theta^l &< \tau_4^l - t_3^l. \label{eq: theta upper bound}
        \end{align}
        And, as in Equation~\eqref{eq: tau2>tau1} to derive Equation~\eqref{eq: theta lower bound} via measurement Equation~\eqref{eq: meas eqn}:
        \begin{align}
            0 &< \tau_2^l - \tau_1^l,  \label{eq: tau2>tau1} \\
            0 &< t_2^l + \theta^l - \tau_1^l, \nonumber \\
            - (t_2^l - \tau_1^l) &< \theta^l. \label{eq: theta lower bound}
        \end{align}
        Combining Equation~\eqref{eq: theta upper bound}~and~\eqref{eq: theta lower bound} yields bounds on $\theta^l$, as in Equation~\eqref{eq: theta bounds}: provably secure against man-in-the-middle attacks.
        Suppose a man-in-the-middle adversary induces delays by increasing $t_2$ or $\tau_4$, then the bounds become looser.
        $\tau_1^l$ is measured directly by the receiver and $t_3^l$ is protected with authentication cryptography in NTS; hence, both have integrity:
        \begin{equation} \label{eq: theta bounds}
            -(t_2^l - \tau_1^l) < \theta^l < \tau_4^l - t_3^l.
        \end{equation}

        The spread between the bounds of $\theta^l$ in Equation~\eqref{eq: theta bounds} is the round-trip time $\tau_4^l - t_3^l + t_2^l - \tau_1^l$, and $\hat \theta$ is the middle of that spread.
        Observing the round-trip time and specifying bounds on $\theta^l$ produce equivalent security on $\theta^l$ in the NTS context; however, the bounds of Equation~\eqref{eq: theta bounds} are more useful in the GNSS TESLA context.

    \subsection{Normal TESLA Loose Time Synchronization} \label{sec: TESLA sync intro}

        {\em Normal} TESLA is a two-way protocol that provides message authentication and integrity~\cite{perrig2005timed}.
        The bulk of messages secured with TESLA utilize a bit-commitment procedure.
        In addition, TESLA utilizes typical asymmetric authentication protocols sparingly {\em as maintenance} to save message bandwidth.
        
        The bit commitment authentication scheme is summarized as follows.
        As the maintenance step, the provider makes a hiding commitment to a key signed with their asymmetric authentication protocol.
        The commitment is the final-derived key in a chain of hidden preimage keys.
        To authenticate messages, the provider will utilize the hidden preimage keys in the reversed derivation order.
        The provider sends a message authentication code generated with a hidden preimage key.
        At a delay $\Theta$ known to the receiver, the sender ceases using that hidden preimage key and reveals the preimage key.
        The provider can repeat until all the hidden preimage keys are utilized.
        Generally, the hiding commitment function and message authentication code is a salted, collision-resistant hash function (e.g., SHA256, HMAC-SHA256).

        To break authentication security, the adversary must do one of the following.
        They must break the authentication security of the asymmetric authentication protocol signing the final-derived key.
        They must break the hiding commitment function's preimage resistance.
        They must break the message authentication code function.
        Or, and the subject of this work, they must fool the receiver into accepting a message authentication code after the release of the corresponding preimage key.
        Once a preimage key is released, then the adversary can use it to generate an arbitrary forgery; hence, the receiver must know to reject messages and message authentication codes after the preimage key release.
        To account for this timing situation, {\em normal} TESLA requires Loose-time Synchronization.

        Loosely time synchronized means the receiver's clock lag to the provider clock is less than a certain bound, which is called $\Theta$ in this work.
        $\Theta$ is the {\em shortest} time between broadcast of {\em any} commitment's last bit (or the last bit of the message if the message comes after the commitment) and the release of the first bit of the associated delay-released key.
        When the receiver clock lags beyond $\Theta$, then the purported security level decreases until the receiver will accept arbitrary forgeries.

        The description of TESLA provides a simple protocol to bootstrap loose-time synchronization \cite{perrig2005timed}, which includes the first half of Algorithm~\ref{alg: nts}.
        In a {\em normal} (i.e., two-way) TESLA-enabled system, the two individual communicating parties can set a safe key-disclosure delay with Equation~\eqref{eq: TESLA bootstrap orig}, where $\Delta t$ is the length of the communication session and $B(\Delta t)$ is the maximum clock drift possible during the session:
        \begin{equation} \label{eq: TESLA bootstrap orig}
            \Theta~=~t_2~-~\tau_1~+~B(\Delta t).
        \end{equation}
        This bootstrap is safe because if a man-in-the-middle increases $t_2$, $\Theta$ increases to accommodate.

    \subsection{Loose Time Synchronization for GNSS}

        In the GNSS context, $\Theta$ is immutably fixed constellation-wide, motivating the methods of this work to accommodate.
        A one-way signal cannot achieve secure time-synchronization~\cite{securetimetransfer}.
        Therefore, the time check {\em must} come from a GIC.
        This means that every TESLA-enabled GNSS receiver must have a GIC that is two-way synchronized externally without GNSS.
        If the receiver is disconnected, the GNSS-independent clock must have a known and bounded drift rate to accommodate security between the networked synchronizations.

        In this work, we modify the TESLA bootstrap procedure to accommodate the broadcast-only context and address a resultant vulnerability (see Section~\ref{sec: NTS changes}).
        TESLA already provides cryptographic authentication security via its construction.
        For this work, proof of authentication security must show that messages and commitments received after the release of the corresponding preimage key will be rejected.
        This method's applicability is principally for upcoming, practical broadcast-only TESLA, whereas previous literature includes a generalized approach~\cite{securetimetransfer}.
        This work aggregates each component necessary for safe synchronization in this context, which includes consolidating and improving upon the results from ~\cite{anderson2023addressing,timesyncmod}.
        \cite{fernandez2020independent} discusses methods and requirements in between clock synchronization and provides the clock synchronization condition but does not provide the needed check on each message and proof of security.
        Furthermore, we discuss the interplay among multiple TESLA instances with different synchronization and specify conditions to ensure authentication security.

        There is a wide breadth of attacks on GNSS receivers~\cite{Psiaki2016}.
        Using public GNSS specifications, a spoofer could generate a consistent signal with a radio.
        When cryptographic signatures are incorporated into a GNSS signal, a more advanced GNSS spoofer could observe the cryptographic information in the signal and then repeat it to a receiver.
        These attacks are broadly categorized as repeater or meaconer attacks.
        Due to the GNSS repeater and meaconer threat on the GNSS-deduced time, a receiver must track its compliance with a GIC.
        The GIC will never approach the accuracy of the time deduced from GNSS; however, it needs to be accurate enough to determine whether the receiver's clock lag is less than a function of $\Theta$ (i.e., $\frac \Theta 2$).
        If the GIC lags provider time more than the coarse check, TESLA's authentication security breaks without any warning to the receiver.
        This results from a receiver accepting a message {\em after} the broadcast of the corresponding delay-released key.
        If the GIC leads provider time too much, the receiver will reject authentic messages, causing false alarms to the user.

    \subsection{Proving Receipt Safety under Adversary Model}

        The TESLA security argument derives from three steps.
        The message authentication code is consistent with the message and delay-released key ({\em message integrity}).
        The delay released key is the correct preimage that derives a key signed via asymmetric cryptography ({\em message authentication}).
        The message and message authentication code arrived before the release of the corresponding key ({\em receipt safety}).
        To break TESLA security, the adversary must break either message integrity, message authentication, or receipt safety.
        To break message integrity or message authentication, the adversary must break the underlying cryptographic protocols (e.g., HMAC or SHA256), which is outside the scope of this work.
        This work focuses on the receipt safety part of TESLA.

        Receipt safety means that it is safe to use the GIC to perform dispositive checks that determine whether to accept or reject a message to enforce that a message and its message authentication code arrived before the release of the corresponding TESLA key.
        This work is necessary to ensure receipt safety because the protocol is broadcast-only and cannot adapt to user-specific key release delays.
        Instead, the user is responsible for maintaining a compliant GIC and using it correctly to accept or reject messages.

        Our adversary model has the following capabilities and limitations.
        The adversary has all the capabilities of a Dolev-Yao adversary, meaning they can overhear, intercept, and create messages but cannot break the underlying cryptographic primitives~\cite{DolevYao,beyondDolevModel}.
        Succinctly, the adversary carries the message.
        The adversary is perfectly synchronized to GNSS time.
        The adversary cannot tamper with the GIC.
        The clock output is strictly positive linear (except when adjusted during synchronization) but may have an offset.
        A non-negative, strictly increasing function bounds the maximum clock-offset growth over time during clock operation.
        When attempting to fool the receiver, the adversary may have zero latency but cannot break the message integrity or authentication of the underlying cryptographic primitives with an efficient algorithm.
        In an extreme case, this could mean that the adversary, receiver, and GNSS satellites are adjacent in orbit.
        And if a modern classical computer can perform a computation in polynomial time, the adversary can do that computation instantaneously. 

        To fool a receiver, the adversary must induce a delay in the signal at least the length of the key delay time $\Theta$.
        After that delay, the adversary can instantaneously compute a forged message and transmit it to the receiver.
        The adversary wins and breaks receipt safety if it can fool a receiver into accepting a message after {\em release} of the delayed key.
        The challenge here is constructing a protocol that ensures receiver safety even when the GIC has an offset.

        To prove that a protocol will ensure receipt safety, one must show that in the presence of our delay-capable adversary, the adversary still cannot fool the protocol into breaking receipt safety.
        In this work, the mathematical arguments will follow a consistent structure.
        We introduce delays induced by the adversary at their election.
        These delays must be non-negative, or otherwise, the delays are not physically achievable.
        Sometimes, we may assume a condition on the offset of the GIC (or otherwise, it is assured by another safe procedure).
        Then, we show that the protocol requires the adversary to induce a negative delay to break receipt safety.
        Since negative delays are impossible, the protocol ensures receipt safety.

        In the GNSS context, ranging signal delays on the order of nanoseconds can manipulate the positioning and timing estimate provided by GNSS. 
        These short delay attacks on the ranging signals are not covered by this work unless these manipulated ranging signals arrive with a delay larger than what would be detectable with Equation~\eqref{eq: th assert condition}. 
        Instead, this work focuses on the GNSS data authentication within the TESLA framework, where the adversary must delay the signal by at least $\Theta$ to generate a message forgery that would be detected under Receipt Safety.

    \subsection{GIC Drift Model} \label{sec: gic drift model}
     
        Suppose that the GIC's clock offset over time is $\theta(t)$ and the clock offset at the last synchronization was $\theta^l$.
        For this work, there exists a non-negative, strictly increasing function $B(\cdot)$ that bounds the maximum GIC clock-offset growth over time during clock operation:
        \begin{equation} \label{eq: bounding condition}
            |\theta(t) - \theta^l| < B(t - t^l).
        \end{equation}

        Splitting Equation~\eqref{eq: bounding condition} into the leading and lagging bounds and substituting Equation~\eqref{eq: theta bounds} yields
        \begin{align}
            - B(t - t^l) &< \theta(t) - \theta^l < B(t - t^l) \nonumber \\
            - B(t - t^l) + \theta^l &< \theta(t) <  \theta^l + B(t - t^l) \nonumber \\
            - B(t - t^l) -(t_2^l - \tau_1^l) &< \theta(t) <  \tau_4^l-t_3^l + B(t - t^l). \label{eq: clock bound over time}
        \end{align}
        These bounds that are a function of time will be abbreviated as
        \begin{equation}
           - B_l < \theta < B_u .
        \end{equation}
        
        In Equation~\eqref{eq: clock bound over time}, the uncertainty on $\theta$ bound by $B_l$ and $B_u$ will expand as time progresses since the last synchronization.
        For the proofs of this work, $B$, and therefore $B_l$ and $B_u$, are absolute bounds.
        The GIC's clock offset will never cross them, which is a non-physical assumption.
        We make this assumption for mathematical conciseness, but these proofs could be extended to accommodate a clock whose offset would not exceed the bounds with high probability, noting the corresponding decrease in security by that probability.

\section{Safe Use of a GIC} \label{sec: clock check}

    Due to the broadcast-only nature of GNSS, receivers must use a GIC to enforce a {\em modified} loose-time synchronization assumption.
    The receiver will record the receipt time of a message and commitment and then await for the delay-released key.
    Based on the TESLA scheme's authenticated metadata and GIC, a receiver should be able to associate messages, commitments, and keys together~\cite{AndersonnJournal}.
    Figure~\ref{fig:cadence} depicts the message-commitment-key cadence between the provider and receiver.
    As the adversary achieves zero latency, $\varepsilon \rightarrow 0$ and Figure~\ref{fig:cadence} would depict vertical transmission arrows for the case of instantaneous message transmission, representing the most conservative case in our future bound derivations.
    Moreover, $\Theta = \min (t_k-t_h, t_k-t_m)$ is a constant for all tuples of a message, commitment, and key.

    \def\lineLength{6}
    \def\eplength{0.6}
    \def\eventlength{1.2}
    \usetikzlibrary{arrows.meta}
    \tikzstyle{connect} = [inner sep=0pt, outer sep=0pt, circle, fill=black, minimum size=0pt]
    \begin{figure}
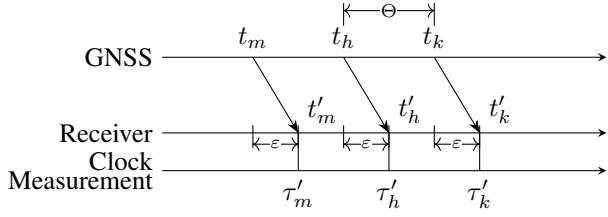

        \centering
        \tikz{
            \node[anchor=east] (GNSS) at (0, 0) {GNSS};
            \node[anchor=east] (Rec) at (0, -1) {Receiver};
            \node[anchor=east, text width=2cm, align=right, font=\linespread{0.5}\selectfont] (RecC) at (0, -1.5) {Clock Measurement};

            \node (GNSSe) at (\lineLength, 0) {};
            \node (Rece) at (\lineLength, -1) {};
            \node (RecCe) at (\lineLength, -1.5) {};

            \draw[-stealth] (GNSS) -- (GNSSe);
            \draw[-stealth] (Rec) -- (Rece);
            \draw[-stealth] (RecC) -- (RecCe);

            \node[connect] (tk) at (4*\eventlength, 0) {};
            \node[connect] (tm) at (\eventlength, 0) {};
            \node[connect] (tmp) at (\eventlength + \eplength, -1) {};
            \node[connect] (taump) at (\eventlength + \eplength, -1.5) {};
            \node[connect] (th) at (2*\eventlength, 0) {};
            \node[connect] (thp) at (2*\eventlength + \eplength, -1) {};
            \node[connect] (tauhp) at (2*\eventlength + \eplength, -1.5) {};
            \node[connect] (tk) at (3*\eventlength, 0) {};
            \node[connect] (tkp) at (3*\eventlength + \eplength, -1) {};
            \node[connect] (taukp) at (3*\eventlength + \eplength, -1.5) {};

            \node[above] (thl) at (th) {$t_h$};
            \node[above] (tml) at (tm) {$t_m$};
            \node[above] (tkl) at (tk) {$t_k$};
            \node[above right] (tmpl) at (tmp) {$t_m'$};
            \node[above right] (tkpl) at (tkp) {$t_k'$};
            \node[above right] (thpl) at (thp) {$t_h'$};
            \node[below] (tauhpl) at (tauhp) {$\tau_h'$};
            \node[below] (taumpl) at (taump) {$\tau_m'$};
            \node[below] (taukpl) at (taukp) {$\tau_k'$};

            \draw[-{Stealth[scale=1]}] (tm) -- (tmp);
            \draw[-{Stealth[scale=1]}] (th) -- (thp);
            \draw[-{Stealth[scale=1]}] (tk) -- (tkp);

            \draw[-] (tmp) -- (taump);
            \draw[-] (thp) -- (tauhp);
            \draw[-] (tkp) -- (taukp);

            \foreach \i in {1, 2, 3} {
                \draw[-] ([yshift=0.1cm]\i * \eventlength, -1) -- ([yshift=-0.2cm]\i * \eventlength, -1);
                \draw[-] ([yshift=0.1cm]\i * \eventlength + \eplength, -1) -- ([yshift=-0.2cm]\i * \eventlength + \eplength, -1);
                \draw[<->] ([yshift=-0.15cm]\i * \eventlength, -1) -- node[midway, fill=white, inner sep=0pt] {\scriptsize$\varepsilon$} ([yshift=-0.15cm]\i * \eventlength + \eplength, -1);
            }

            \draw[-] ([yshift=0.2cm]thl.north) -- ([yshift=-0.1cm]thl.north);
            \draw[-] ([yshift=0.2cm]tkl.north) -- ([yshift=-0.1cm]tkl.north);
            \draw[<->] ([yshift=0.1cm]thl.north) -- node[midway, fill=white, inner sep=0pt] {\scriptsize$\Theta$} ([yshift=0.1cm]tkl.north);
        }
        \caption{
        A conceptual diagram of TESLA's message, commitment (e.g., HMAC), delay-release key transmission cadence from provider to receiver (m, h, k in the diagram, respectively). 
        The transmission arrows approach vertical for the conservative case with a zero-latency adversary as $\varepsilon \rightarrow 0$, which aids in the conciseness of the proofs provided in Section~\ref{sec: clock check}. 
        The fixed commitment delay is set constellation-wide as $\Theta = \min (t_k-t_h, t_k-t_m)$ among all tuples, determining the requirements on receiver GICs.
        }
        \label{fig:cadence}
    \end{figure}

    Safe use of the receiver clock requires that it be GNSS-independent.
    It should be modified using the procedure described in Section~\ref{sec: sync}, and never changed with broadcast-only GNSS.
    To prohibit the acceptance of forgeries, the receiver must ensure two conditions (noting Section~\ref{sec: alt}), which are incorporated in Algorithm~\ref{alg: mhk receipt}.
    First, the GIC must synchronized such that
    \begin{equation}\label{eq: sync cond}
        -\frac \Theta 2 < -B_l < \theta.
    \end{equation}
    Second, the GIC must assert that each message, commitment, and key tuple satisfies
    \begin{equation} \label{eq: th assert condition}
        \max (\tau_m', \tau_h') < t_k - B_l.
    \end{equation}
    The conditions of Equation~\eqref{eq: sync cond}~and~\eqref{eq: th assert condition} are sufficient to have receipt safety.

    Equation~\eqref{eq: th assert condition} is intended to be a broad condition that accommodates many TESLA-enabled GNSS system designs.
    For instance, $\max (\tau_m', \tau_h')$ is the latest release time of any piece of information corresponding to $k$ when there are multiple messages and commitments associated with a key.
    In the case that the system has a simple cadence like Figure~\ref{fig:cadence}, then the condition simplifies.

    We note that in Equation~\eqref{eq: th assert condition}, $\tau_m'$ and $\tau_h'$ are measurements with the GIC that meets the condition of Equation~\eqref{eq: sync cond}.
    Whereas, $t_k$ is determined by the schedule that the GNSS Provider adheres to.

    After determining a message has receipt safety, the receiver must perform additional cryptographic TESLA checks to ensure message integrity and authentication.

    \subsection{Proof of Receipt Safety}

        In this section, we show that Equation~\eqref{eq: sync cond}~and~\eqref{eq: th assert condition} are sufficient to have receipt safety.
        When the message and commitment are received before the release of the key, the message and commitment are sufficient for receipt safety.

        \usetikzlibrary{arrows.meta}
        \begin{figure}
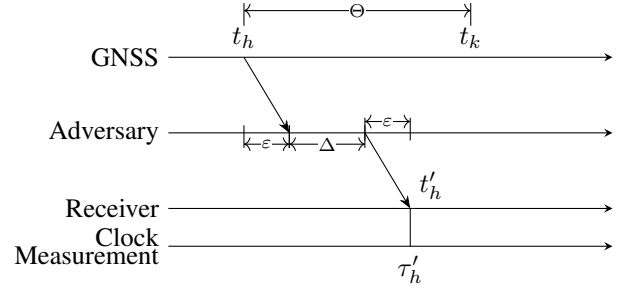

            \def\lineLength{6}
            \def\eplength{0.6}
            \def\eventlength{1}
            \tikzstyle{connect} = [inner sep=0pt, outer sep=0pt, circle, fill=black, minimum size=0pt]
            \tikz{
                \node[anchor=east] (GNSS) at (0, 0) {GNSS};
                \node[anchor=east] (Adv) at (0, -1) {Adversary};
                \node[anchor=east] (Rec) at (0, -2) {Receiver};
                \node[anchor=east, text width=2cm, align=right, font=\linespread{0.5}\selectfont] (RecC) at (0, -2.5) {Clock Measurement};

                \node (GNSSe) at (\lineLength, 0) {};
                \node (Adve) at (\lineLength, -1) {};
                \node (Rece) at (\lineLength, -2) {};
                \node (RecCe) at (\lineLength, -2.5) {};

                \draw[-stealth] (GNSS) -- (GNSSe);
                \draw[-stealth] (Adv) -- (Adve);
                \draw[-stealth] (Rec) -- (Rece);
                \draw[-stealth] (RecC) -- (RecCe);

                \node[connect] (tk) at (4*\eventlength, 0) {};
                \node[connect] (th) at (\eventlength, 0) {};
                \node[connect] (tha) at (\eventlength + \eplength, -1) {};
                \node[connect] (thb) at (2*\eventlength + \eplength, -1) {};
                \node[connect] (thp) at (2*\eventlength + 2*\eplength, -2) {};
                \node[connect] (tauhp) at (2*\eventlength + 2*\eplength, -2.5) {};
                \draw[-{Stealth[scale=1]}] (th) -- (tha);
                \draw[-{Stealth[scale=1]}] (thb) -- (thp);
                \draw[-] (thp) -- (tauhp);

                \node[above] (thl) at (th) {$t_h$};
                \node[above right] at (thp) {$t_h'$};
                \node[below] at (tauhp) {$\tau_h'$};
                \node[above] (tkl) at (tk) {$t_k$};

                \draw[-] ([yshift=0.1cm]\eventlength, -1) -- ([yshift=-0.2cm]\eventlength, -1);
                \draw[-] ([yshift=0.1cm]\eventlength + \eplength, -1) -- ([yshift=-0.2cm]\eventlength + \eplength, -1);
                \draw[<->] ([yshift=-0.15cm]\eventlength, -1) -- node[midway, fill=white, inner sep=0pt] {\scriptsize$\varepsilon$} ([yshift=-0.15cm]\eventlength + \eplength, -1);

                \draw[-] ([yshift=0.1cm]\eventlength + \eplength, -1) -- ([yshift=-0.2cm]\eventlength + \eplength, -1);
                \draw[-] ([yshift=0.1cm]2*\eventlength + \eplength, -1) -- ([yshift=-0.2cm]2*\eventlength + \eplength, -1);
                \draw[<->] ([yshift=-0.15cm]\eventlength + \eplength, -1) -- node[midway, fill=white, inner sep=0pt] {\scriptsize$\Delta$} ([yshift=-0.15cm]2*\eventlength + \eplength, -1);

                \draw[-] ([yshift=0.2cm]2*\eventlength + \eplength, -1) -- ([yshift=-0.1cm]2*\eventlength + \eplength, -1);
                \draw[-] ([yshift=0.2cm]2*\eventlength + 2*\eplength, -1) -- ([yshift=-0.1cm]2*\eventlength + 2*\eplength, -1);
                \draw[<->] ([yshift=0.15cm]2*\eventlength + \eplength, -1) -- node[midway, fill=white, inner sep=0pt] {\scriptsize$\varepsilon$} ([yshift=0.15cm]2*\eventlength + 2*\eplength, -1);

                \draw[-] ([yshift=0.2cm]thl.north) -- ([yshift=-0.1cm]thl.north);
                \draw[-] ([yshift=0.2cm]tkl.north) -- ([yshift=-0.1cm]tkl.north);
                \draw[<->] ([yshift=0.1cm]thl.north) -- node[midway, fill=white, inner sep=0pt] {\scriptsize$\Theta$} ([yshift=0.1cm]tkl.north);
            }
            \caption{
                A conceptual diagram of the spoofing scenario as a GNSS message traverses to the receiver.
                GNSS transmits commitment $h$, which the adversary delivers to the receiver with an adversary-selected delay $\Delta$.
                The adversary and scenario have latencies $\varepsilon$.
                We allow $\varepsilon \rightarrow 0$ for the sake of proving the worst case.
                Commitment $h$ is transmitted and received at time $t_h$ and $t_h'$, respectively, in the clock frame provided by GNSS.
                Commitment $h$ is measured to have been received at $\tau_h'$ under a clock offset of $\theta$.
                If the receiver accepts $h$ after GNSS releases $k$, then the receiver would accept an arbitrary forgery, which occurs when $\Delta \geq \Theta$.
                $\Theta$ is the system-wide minimum time spread between the transmission of the last bit of $h$ and the first bit of $k$.
                }
            \label{fig: time sync proof diagram}
        \end{figure}

        As conceptually diagrammed in Figure~\ref{fig: time sync proof diagram}, GNSS releases the commitment $h$ at $t_h$, where the adversary observes, modifies, and then delivers it to the receiver.
        Commitment $h$ is received by the receiver at $t_h'$, which is measured by the receiver GIC as time $\tau_h'$.
        Like Figure~\ref{fig:cadence}, Figure~\ref{fig: time sync proof diagram} assumes that the commitment $h$ is the latest released object associated with $k$ without loss of generality.

        We let the adversary have zero latency, so each $\varepsilon \rightarrow 0$, and the adversary is perfectly synchronized with GNSS.
        In Figure~\ref{fig: time sync proof diagram}, the adversary wins if the receiver accepts a commitment $h$ after GNSS releases the corresponding key $k$.
        This corresponds to when $\Delta \geq \Theta$, when the adversary induces sufficient delay that it can forge an arbitrary message and commitment with knowledge of the released $k$.

        Now, we will prove that if a receiver certifies that its GIC synchronization meets the condition of Equation~\eqref{eq: sync cond} and the enforces condition of Equation~\eqref{eq: th assert condition}, then the the message and its commitment arrived before the release of the key (i.e., $\max (t_m', t_h') < t_k$).
        And therefore, the message has receipt safety.
        
        We start with the enforced condition
        \begin{align}
            \max (\tau_m', \tau_h') &< t_k - B_l. \tag{\ref{eq: th assert condition}}
        \end{align}
        Then, by substituting the measurement equation, we have
        \begin{align}
            \max (t_m', t_h') + \theta &< t_k - B_l, \\
            \theta &< t_k - B_l - \max (t_m', t_h').
        \end{align}
        Next, we substitute the presumed clock synchronization condition and simplify with
        \begin{align}
            -B_l &< t_k - B_l - \max (t_m', t_h'), \\
            \max (t_m', t_h') &< t_k.
        \end{align}
        Since the message and commitments were received {\em in the GNSS clock frame} before the release of the corresponding key, it has receipt safety under TESLA.

    \subsection{Alternative Condition Design}\label{sec: alt}

        \definecolor{cardinalred}{RGB}{140, 21, 21}
        \definecolor{skydark}{HTML}{016895}
        \definecolor{olivedark}{HTML}{7A863B}
        \colorlet{thesisaccentwarn}{cardinalred}
        \colorlet{thesiscolor1}{skydark}
        \colorlet{thesiscolor2}{olivedark}

        \tikzstyle{pt} =  [inner sep=0pt, outer sep=0pt, minimum size=0pt]
        \begin{figure}[tbp]
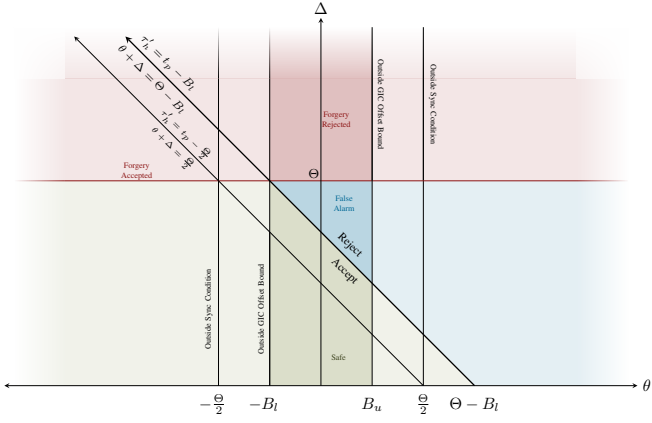

        \centering
        \resizebox{\linewidth}{!}{
        \def\thetalength{1.1}
        \tikz{
            \foreach \i in {-6, -5, -4, -3, -2, -1, 0, 1, 2, 3, 4, 5, 6}{
                \foreach \j in {0, 1, 2, 3, 4, 5, 6, 7}{
                    \node[pt] (p\i\j) at (\i*\thetalength, \j*\thetalength) {};
                }
            } 

            \filldraw[fill=thesiscolor2!30, draw=none, opacity=0.90] (p-10.center) -- (p10.center) -- (p12.center) -- (p-14.center) -- cycle;
            \filldraw[fill=thesiscolor2!30, draw=none, opacity=0.35] (p-50.center) -- (p-10.center) -- (p-14.center) -- (p-54.center) -- cycle;
            \filldraw[fill=thesiscolor2!30, draw=none, opacity=0.35] (p10.center) -- (p30.center) -- (p12.center) -- cycle;
            \filldraw[fill=thesiscolor2!30, draw=none, opacity=0.35, path fading=west] (p-60.center) -- (p-50.center) -- (p-54.center) -- (p-64.center) -- cycle;

            \filldraw[fill=thesiscolor1!30, draw=none, opacity=0.90] (p12.center) -- (p14.center) -- (p-14.center) -- cycle;
            \filldraw[fill=thesiscolor1!30, draw=none, opacity=0.35] (p30.center) -- (p50.center) -- (p54.center) -- (p14.center) -- (p12.center) -- cycle;
            \filldraw[fill=thesiscolor1!30, draw=none, opacity=0.35, path fading=east] (p50.center) -- (p60.center) -- (p64.center) -- (p54.center) -- cycle;

            \filldraw[fill=thesisaccentwarn!30, draw=none, opacity=0.90] (p-14.center) rectangle (p16.center);
            \filldraw[fill=thesisaccentwarn!30, draw=none, opacity=0.35] (p14.center) rectangle (p56.center);
            \filldraw[fill=thesisaccentwarn!30, draw=none, opacity=0.35] (p-54.center) rectangle (p-16.center);
            \filldraw[fill=thesisaccentwarn!30, draw=none, opacity=0.35, path fading=east] (p54.center) -- (p64.center) -- (p66.center) -- (p56.center) -- cycle;
            \filldraw[fill=thesisaccentwarn!30, draw=none, opacity=0.35, path fading=west] (p-54.center) -- (p-64.center) -- (p-66.center) -- (p-56.center) -- cycle;
            \filldraw[fill=thesisaccentwarn!30, draw=none, opacity=0.90, path fading=north] (p-16.center) -- (p16.center) -- (p17.center) -- (p-17.center) -- cycle;
            \filldraw[fill=thesisaccentwarn!30, draw=none, opacity=0.35, path fading=north] (p16.center) -- (p56.center) -- (p57.center) -- (p17.center) -- cycle;
            \filldraw[fill=thesisaccentwarn!30, draw=none, opacity=0.35, path fading=north] (p-16.center) -- (p-56.center) -- (p-57.center) -- (p-17.center) -- cycle;
            \shade[lower right=thesisaccentwarn!5, upper left=thesisaccentwarn!0, draw=none, opacity=1] (p-56.center) -- (p-66.center) -- (p-67.center) -- (p-57.center) -- cycle;
            \shade[lower left=thesisaccentwarn!5, upper right=thesisaccentwarn!0, draw=none, opacity=1] (p56.center) -- (p66.center) -- (p67.center) -- (p57.center) -- cycle;

            \node at ([yshift=-0.4cm]p20.center){$\frac \Theta 2$};
            \node at ([xshift=-0.1275cm, yshift=-0.4cm]p-20.center){$-\frac \Theta 2$};

            \draw[-] (p14) -- node[above, sloped, rotate=180] {\tiny Outside GIC Offset Bound} (p17);
            \draw[-] (p-10) -- node[above, sloped] {\tiny Outside GIC Offset Bound} (p-13);

            \draw[-] (p24) -- node[above, sloped, rotate=180] {\tiny Outside Sync Condition} (p27);
            \draw[-] (p-20) -- node[above, sloped] {\tiny Outside Sync Condition} (p-23);

            \draw[-] (p20) -- (p26);
            \draw[-, path fading=north] (p26) -- (p27);
            \draw[-] (p-20) -- (p-26);
            \draw[-, path fading=north] (p-26) -- (p-27);

            \draw[-, thesisaccentwarn] (p-54) -- (p54);
            \draw[-, thesisaccentwarn, path fading=west] (p-64) -- (p-54);
            \draw[-, thesisaccentwarn, path fading=east] (p54) -- (p64);

            \node[thesisaccentwarn, anchor=south west, xshift=-0.1cm, yshift=-0.1cm, align=center, font=\tiny] at (p05) {Forgery \\ Rejected};
            \node[thesiscolor1, anchor=north west, xshift=0.15cm, yshift=-0.2cm, font=\tiny, align=center] at (p04) {False \\ Alarm};
            \node[thesisaccentwarn, anchor=south east, xshift=-0.2cm, yshift=-0.1cm, align=center, font=\tiny] at ([xshift=-0.0cm]p-34) {Forgery \\ Accepted};
            \node[thesiscolor2!50!black, anchor=west, font=\tiny, align=center] at ([xshift=0.1cm, yshift=-0.5cm]p01) {Safe};
           
            \draw[stealth-] ([xshift=0.2cm, yshift=-0.2cm]p-57.center) -- node[pos=0.30, rotate=-45, yshift=0.2cm] {\tiny$\tau_h' = t_p - \frac \Theta 2$} node[pos=0.30, rotate=-45, yshift=-0.2cm] {\tiny$\theta + \Delta = \frac \Theta 2$} (p20.center);

            \draw[stealth-,thick] ([xshift=0.2cm, yshift=-0.2cm]p-47.center)  -- node[pos=0.10, rotate=-45, yshift=0.25cm] {\scriptsize$\tau_h' = t_p - B_l$} node[pos=0.10, rotate=-45, yshift=-0.25cm] {\scriptsize$\theta + \Delta = \Theta - B_l$} node[above, pos=0.62, rotate=-45] {\scriptsize Reject} node[below, pos=0.65, rotate=-45] {\scriptsize Accept} (p30.center);

            \draw[-] (p10) -- (p16);
            \draw[-, path fading=north] (p16) -- (p17);
            \draw[-] (p-10) -- (p-16);
            \draw[-, path fading=north] (p-16) -- (p-17);

            \node at ([yshift=-0.4cm]p10.center){$B_u$};
            \node at ([xshift=-0.1275cm, yshift=-0.4cm]p-10.center){$-B_l$};

            \node at ([xshift=-0cm, yshift=-0.4cm]p30.center){$\Theta -B_l$};

            \draw[stealth-stealth] ([xshift=-0.2cm]p-60.center) -- ([xshift=0.2cm]p60.center);
            \node at ([xshift=0.4cm]p60.center){$\theta$};
            \draw[-stealth] (p00.center) -- ([yshift=0.2cm]p07.center);

            \node at ([yshift=0.4cm]p07.center){$\Delta$};
            \node at ([xshift=-0.15cm, yshift=0.15cm]p04.center){\footnotesize$\Theta$};
        }
        }
        \caption{
            Conceptual Diagram Depicting Safety Regions Among Adversary-selected Delay And GIC Clock State.
            Among the the adversary-selected state $\Delta$ and the GIC drift state $\theta$, there must be constraints to prohibit forgery.
            The figure shows the safety condition line and labels regions of interest.
            While our selected decision line is shown, the GNSS designer can shift the regions left or right.
            Here,  the uncertainty of the clock drift is evenly spread to accommodate unbiased clock drift in the context of false alarms.
            The entire region Forgery Accepted is outside the time synchronization bound.
            As time progresses, the decision the decision line will move down, expanding the False Alarm region until the GIC must synchronize.
            }
        \label{fig: time sync design}
        \end{figure}

        From Equation~\eqref{eq: sync cond}, the receipt safety conditions relate to $\frac \Theta 2$.
        But what about beyond check $\max (\tau_m', \tau_h') < t_p - B_l$ and synchronization condition $\theta > -\frac \Theta 2$?
        And why $\frac \Theta 2$, as opposed to $\Theta$?

        From an intuition point of view, the $\frac \Theta 2$ comes from the following scenario.
        Suppose that a receiver's clock is $\frac \Theta 2$ behind.
        If an adversary delays the messages by $\Theta$, it (1) can forge messages, and (2) the message will induce the receiver to believe its clock is $\frac \Theta 2$ {\em ahead}.
        $\frac \Theta 2$ ahead is also within the safe condition.
        Hence, the entire safe region cannot be larger than $\Theta$, and $\frac \Theta 2$ spreads the uncertainty in both directions evenly for unbiased clock drift.

        Without loss of generality, lets suppose that the GNSS system has single-object cadence: $m$, $h$, and $k$.
        Therefore, the decision boundary becomes $\tau_h' = t_k - B_l$.
        We have two independent states in our scenario: $\Delta$ and $\theta$.
        The adversary may elect any $\Delta \geq 0$, and the clock drift $\theta$ may be any value, subject to the clock offset bounds.
        There is no receipt safety when $\Delta \geq \Theta$, where an adversary may forge an arbitrary message and commitment with the released key.
        Substituting decision boundary $\tau_h' = t_k - B_l$ yields functions of the state variables with
        \begin{align}
            \tau_h' = t_k - B_l, \\
            t_h' + \theta = t_k - B_l, \nonumber \\
            t_h + \Delta + \theta = t_k - B_l, \label{eq: contr eq}\\
            \Delta + \theta = t_k-t_h - B_l, \nonumber \\
            \Delta + \theta = \Theta - B_l.
        \end{align}

        From states $\Delta$ and $\theta$, we plot them with Figure~\ref{fig: time sync design} to visualize a receiver's operating condition to assist with design.
        The check must ensure that no adversary-selected $\Delta$ exists under any possible clock condition $\theta$ that would have the receiver accept a forgery.
        Because $\theta$ and $\Delta$ contribute equally to $\tau_h'$ in Equation~\eqref{eq: contr eq}, the slope of the line in Figure~\ref{fig: time sync design} cannot be changed.
        However, one could raise or lower the decision line, provided the synchronization requirement is changed.

        For instance, a receiver could enforce the condition $\tau_h' < t_k$ and require that $\theta > -\Theta$.
        This corresponds to moving each boundary line and shape in the figure to the left by $\frac \Theta 2$.
        The issue with doing this change is the potential for false alarms.
        If we reasonably assume that, at synchronization, $\theta=0$, then depending on the clock drift, the check would have a 50\% chance of false alarm.
        Optimizing this condition may be useful if the receiver knows its clock drift rate is biased, but otherwise, it would be best to use the original condition and proof above.

        Our choice of a decision boundary with $\frac \Theta 2$ spreads the slack evenly.
        For the rest of this work, we adopt this evenly split condition.
        This also means that Equation~\eqref{eq: sync cond} must be amended to bind the leading case to accommodate false alarms to become
        \begin{equation}\label{eq: sync cond both}
            -\frac \Theta 2 < -B_l < \theta < B_u < \frac \Theta 2.
        \end{equation}

    \subsection{Authentication Security Level}

        GNSS systems messages are rigidly scheduled.
        Because of this, a GNSS designer knows that each GNSS message and message authentication code has one chance to be authenticated.
        For instance, at a 50 bps connection for GPS C/A, the security designer need not consider a scenario where the adversary may send numerous perturbed copies of a navigation message in attempts of accepting a single forgery.
        The subject of each message in the rigid scheduled has already been assigned by the signal specification.

        For the security designer, this leads to an important consideration.
        The adversary will only have a single try to spoof an individual message under TESLA-enabled GNSS.
        Suppose each commitment h includes $n_\text{h}$ bits.
        Then, assuming perfect adherence to the security properties of a particular cryptographic primitive (e.g., HMAC-SHA256), the adversarial spoofing probability will be $2^{-n_h}$.
        In expectation, the adversary should expect a successful forgery every $2^{-n_h}$ tries.
        The security is {\em not} subject to birthday attacks because the hash function is not known until the release of the key.
        Therefore, the security provided by this step is $n_h$ bits.

        Suppose each key k includes $n_\text{k}$ bits.
        Before the release of the key, the authentication security of a particular key is $n_k$ bits.
        The birthday attack does not apply because the adversary must defeat preimage resistance.
        Therefore, before release of the key the final message authentication security level is $\min(n_\text{h}, n_\text{k})$ bits.
        All TESLA-enabled GNSS designs have proposed very short HMACs on the order of $n_h=32$ to accommodate the GNSS data bandwidth constraint.
        Then $n_k$ is the typical cryptographic security on the level of 128-bits for the keys themselves.

\section{Safe Synchronization for TESLA-enabled GNSS} \label{sec: safe check and sync}

    In this section, we provide two useful and dispositive procedures for broadcast-only TESLA loose-time synchronization for TESLA-enabled GNSS, and we prove their security under our adversary.
    In Section~\ref{sec: cert}, we describe how to certify the safety of the GIC performing the receipt safety checks without making any adjustments to the clock.
    A receiver must enforce Equations~\eqref{eq: safe false alerts}~and~\eqref{eq: safe security} using an NTS query.
    In Section~\ref{sec: sync}, we describe how to safely synchronize.
    The receiver must enforce Equation~\eqref{eq: theta star bounds} on any adjustment suggested by an NTS query.
    The conditions proven in this section are restated in Algorithm~\ref{alg: correct time sync} for convenience of those who are not concerned with the proofs of safety.

    \subsection{Certifying Clock Safety for Receipt Safety} \label{sec: cert}

        We start with bounding the clock drift over time $\theta(t)$ to the clock drift at the last synchronization $\theta^l$ with the non-negative, strictly increasing bounding function $B(\cdot)$:
        \begin{align}
            |\theta(t) - \theta^l| &< B(t - t^l). \tag{\ref{eq: bounding condition}}
        \end{align}
        There are two relevant derivation cases, given the absolute value.
        For the leading case, we have the following, substituting Equation~\eqref{eq: theta bounds}:
        \begin{align}
            \theta(t) - \theta^l &< B(t - t^l), \\
            \theta(t) - B(t - t^l) &< \theta^l, \nonumber \\
            \theta(t) - B(t - t^l) &<  \tau_4^l - t_3^l, \nonumber \\
            \theta(t) &< \tau_4^l - t_3^l + B(t - t^l).
        \end{align}
        To prevent the false alert, we must certify that $\theta(t) < \frac \Theta 2$, but our safe estimate is $\tau_4^l - t_3^l + B(t - t^l)$.
        Therefore, we desire our measured bound to be tighter than the safety condition on false alerts, arriving at Equation~\eqref{eq: safe false alerts}:
        \begin{equation} \label{eq: safe false alerts}
            \tau_4^l - t_3^l + B(t - t^l) < \frac \Theta 2.
        \end{equation}

        For the lagging case, we follow a similar procedure:
        \begin{align}
            - B(t - t^l) &< \theta(t) - \theta^l, \\
            \theta^l &< \theta(t) + B(t - t^l),  \nonumber \\
            -(t_2^l - \tau_1^l) &< \theta(t) + B(t - t^l),  \nonumber \\
            -(t_2^l - \tau_1^l) - B(t - t^l) &< \theta(t).         
        \end{align}
        To prevent breaking receipt safety, we must certify that $-\frac \Theta 2 < \theta(t)$.
        Therefore, once again, we desire our measured bound to be tighter than the safety condition on security, arriving at Equation~\eqref{eq: safe security}:
        \begin{equation} \label{eq: safe security}
            -\frac \Theta 2 < -(t_2^l - \tau_1^l) - B(t - t^l).
        \end{equation}

        The receiver can use conditions from Equation~\eqref{eq: safe false alerts}~and~\eqref{eq: safe security} to determine the safety of its GIC at the query time (when $B\approx 0$).
        Provided an accurate model of $B$, the receiver can determine the time of the next appropriate synchronization. 
        Since $B$ is non-negative and strictly increasing, the receiver can uniquely solve Equations~\eqref{eq: safe false alerts}~and~\eqref{eq: safe security} for the next appropriate synchronization time.
        That time will be the minimum $t$ between the maximum $t$ satisfying Equation~\eqref{eq: safe false alerts}~and the maximum $t$ satisfying Equation~\eqref{eq: safe security}.

        One could add Equations~\eqref{eq: safe false alerts}~and~\eqref{eq: safe security} to provide a single safety condition as a function of $l$'s round trip time.
        However, the separation is helpful for the security proofs since security relies on the lagging bound.

        \subsubsection{Proof of Receipt Safety}

\begin{figure}
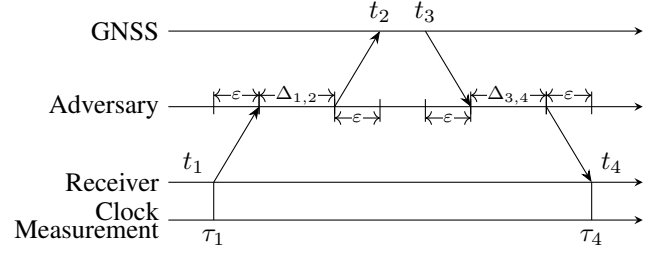

            \tikz{
                \def\lineLength{6.8}
                \def\eplength{0.6}
                \def\eventlength{1}
                \node[anchor=east] (GNSS) at (0.4, 0) {GNSS};
                \node[anchor=east] (Adv) at (0.4, -1) {Adversary};
                \node[anchor=east] (Rec) at (0.4, -2) {Receiver};
                \node[anchor=east, text width=2cm, align=right, font=\linespread{0.5}\selectfont] (RecC) at (0.4, -2.5) {Clock Measurement};

                \node (GNSSe) at (\lineLength, 0) {};
                \node (Adve) at (\lineLength, -1) {};
                \node (Rece) at (\lineLength, -2) {};
                \node (RecCe) at (\lineLength, -2.5) {};

                \draw[-stealth] (GNSS) -- (GNSSe);
                \draw[-stealth] (Adv) -- (Adve);
                \draw[-stealth] (Rec) -- (Rece);
                \draw[-stealth] (RecC) -- (RecCe);

                \node[connect] (tau1p) at (\eventlength, -2.5) {};
                \node[connect] (t1p) at (\eventlength, -2) {};

                \node[below] at (tau1p) {$\tau_1$};
                \node[above left] at (t1p) {$t_1$};
                \draw[-] (tau1p) -- (t1p);

                \node[connect] (tha) at (\eventlength + \eplength, -1) {};
                \draw[-{Stealth[scale=1]}] (t1p) -- (tha);
                \draw[<->] ([yshift=0.15cm]\eventlength, -1) -- node[midway, fill=white, inner sep=0pt] {\scriptsize$\varepsilon$} ([yshift=0.15cm]\eventlength + \eplength, -1);
                \draw[-] ([yshift=0.2cm]\eventlength, -1) -- ([yshift=-0.1cm]\eventlength, -1);
                \draw[-] ([yshift=0.2cm]\eventlength + \eplength, -1) -- ([yshift=-0.1cm]\eventlength + \eplength, -1);

                \node[connect] (thb) at (2*\eventlength + \eplength, -1) {};
                \draw[-] ([yshift=0.2cm]\eventlength + \eplength, -1) -- ([yshift=-0.1cm]\eventlength + \eplength, -1);
                \draw[-] ([yshift=0.2cm]2*\eventlength + \eplength, -1) -- ([yshift=-0.1cm]2*\eventlength + \eplength, -1);
                \draw[<->] ([yshift=0.15cm]\eventlength + \eplength, -1) -- node[midway, fill=white, inner sep=0pt] {\scriptsize$\Delta_{1,2}$} ([yshift=0.15cm]2*\eventlength + \eplength, -1);

                \node[connect] (t2) at (2*\eventlength + 2 * \eplength, 0) {};

                \draw[-{Stealth[scale=1]}] (thb) -- (t2);
                \node[above] (th2) at (t2) {$t_2$};

                \draw[-] ([yshift=0.1cm]2*\eventlength + \eplength, -1) -- ([yshift=-0.2cm]2*\eventlength + \eplength, -1);
                \draw[-] ([yshift=0.1cm]2*\eventlength + 2*\eplength, -1) -- ([yshift=-0.2cm]2*\eventlength + 2*\eplength, -1);
                \draw[<->] ([yshift=-0.15cm]2*\eventlength + \eplength, -1) -- node[midway, fill=white, inner sep=0pt] {\scriptsize$\varepsilon$} ([yshift=-0.15cm]2*\eventlength + 2*\eplength, -1);                

                \node[connect] (t3) at (2*\eventlength + 3 * \eplength, 0) {};
                \node[above] (th3) at (t3) {$t_3$};

                \node[connect] (thc) at (2*\eventlength + 4 * \eplength, -1) {};
                \draw[-{Stealth[scale=1]}] (t3) -- (thc);
                \draw[<->] ([yshift=-0.15cm]2*\eventlength + 3 * \eplength, -1) -- node[midway, fill=white, inner sep=0pt] {\scriptsize$\varepsilon$} ([yshift=-0.15cm]2*\eventlength + 4 * \eplength, -1);
                \draw[-] ([yshift=0.1cm]2*\eventlength + 3 * \eplength, -1) -- ([yshift=-0.2cm]2*\eventlength + 3 * \eplength, -1);
                \draw[-] ([yshift=0.1cm]2*\eventlength + 4 * \eplength, -1) -- ([yshift=-0.2cm]2*\eventlength + 4 * \eplength, -1);

                \node[connect] (thd) at (3*\eventlength + 4 * \eplength, -1) {};
                \draw[-] ([yshift=0.2cm]2*\eventlength + 4 * \eplength, -1) -- ([yshift=-0.1cm]2*\eventlength + 4 * \eplength, -1);
                \draw[-] ([yshift=0.2cm]3*\eventlength + 4 * \eplength, -1) -- ([yshift=-0.1cm]3*\eventlength + 4 * \eplength, -1);
                \draw[<->] ([yshift=0.15cm]2*\eventlength + 4 * \eplength, -1) -- node[midway, fill=white, inner sep=0pt] {\scriptsize$\Delta_{3,4}$} ([yshift=0.15cm]3*\eventlength + 4 * \eplength, -1);

                \draw[<->] ([yshift=0.15cm]3*\eventlength + 4 * \eplength, -1) -- node[midway, fill=white, inner sep=0pt] {\scriptsize$\varepsilon$} ([yshift=0.15cm]3*\eventlength + 5 * \eplength, -1);
                \draw[-] ([yshift=0.2cm]3*\eventlength + 4 * \eplength, -1) -- ([yshift=-0.1cm]3*\eventlength + 4 * \eplength, -1);
                \draw[-] ([yshift=0.2cm]3*\eventlength + 5 * \eplength, -1) -- ([yshift=-0.1cm]3*\eventlength + 5 * \eplength, -1);

                \node[connect] (t4p) at (3*\eventlength + 5 * \eplength, -2) {};
                \draw[-{Stealth[scale=1]}] (thd) -- (t4p);
                
                \node[connect] (tau4p) at (3*\eventlength + 5 * \eplength, -2.5) {};
                \node[below] at (tau4p) {$\tau_4$};
                \node[above right] at (t4p) {$t_4$};
                \draw[-] (tau4p) -- (t4p);

            }
                \caption{
                    Conceptual diagram of the attack scenario during the GIC check and synchronization scenario.
                    The adversary may induce delays $\Delta_{1,2}, \Delta_{3,4} \geq 0$ and have zero latency (i.e., $\varepsilon \rightarrow 0$).
                    The event numbers come from the NTS protocol.
                }
                \label{fig: cert attack}
            \end{figure}

            To show clock safety in the receipt safety check, we will show that if an adversary delays messages in the protocol, the adversary cannot fool a receiver into believing its unsafe clock is safe.
            The relevant condition for this proof is Equation~\eqref{eq: safe security}.
            The other condition that helps with false alarms is not relevant to safety in this section.
            The attack is conceptually diagrammed with Figure~\ref{fig: cert attack}.

            In Figure~\ref{fig: cert attack}, the zero-latency (i.e., $\varepsilon \rightarrow 0$) adversary may elect to introduce a delay $\Delta_{1,2} \geq 0$.
            Now suppose that the receiver's clock is broken, meaning $\theta^l < -\frac \Theta 2$.
            The adversary must select a $\Delta_{1,2} \geq 0$ that passes Equation~\eqref{eq: safe security} to fool the receiver into believing its clock is safe.

            Starting from the check with Equation~\eqref{eq: safe security}, we substitute the measurement equation:
            \begin{align}
                -\frac \Theta 2 &< -(t_2^l - \tau_1^l) - B(t - t^l), \tag{\ref{eq: safe security}} \\
                -\frac \Theta 2 &< -(t_2^l - t_1^l - \theta^l) - B(t - t^l), \nonumber \\
                &-\frac \Theta 2 + t_2^l - t_1^l + B(t - t^l) < \theta^l. \nonumber 
            \end{align}

            We substitute the adversarial delay and apply the broken-clock condition:
            \begin{align}
                -\frac \Theta 2 + \Delta_{1,2} + B(t - t^l) &< \theta^l,  \nonumber \\
                -\frac \Theta 2 + \Delta_{1,2} + B(t - t^l) &< - \frac \Theta 2, \nonumber  \\
                \Delta_{1,2} + B(t - t^l) &< 0.
            \end{align}

            Both $\Delta_{1,2}, B(\cdot) \geq 0$.
            Being generous to the adversary, we let $B(\cdot) = 0$.
            Therefore, 
            \begin{align}
                \Delta_{1,2} &< 0.
            \end{align}
            This contradicts the initial requirement that the adversary induces $\Delta_{1,2}\geq 0$; therefore, the adversary cannot fool the receiver into believing its broken clock is safe.

            From an intuition point of view, when the adversary induces $\Delta_{1,2}$, the adversary causes the receiver to believe its clock lag is worse.
            As the receiver's lagging clock belief gets worse, the check condition becomes more brittle.
            This means that the adversary's delay will make the alert trigger more sensitive.
            This argument also applies to $B(\cdot)$.

            Suppose an adversary were to induce a $\Delta_{3,4} \geq 0$ on the NTS query return.
            This manipulation would decrease the upper bound on $\theta$ in Equation~\eqref{eq: safe false alerts}.
            But since $\theta$ still adheres to the lower bound of Equation~\eqref{eq: safe security}, the upper bound is not the active constraint on $\theta$ in the proof of detecting an unsafe clock with the worst $\theta$ under adversarial manipulation.

    \subsection{Synchronizing Receiver Clock Safely} \label{sec: sync}

        Suppose that the receiver uses an NTS query to compute a $\delta\theta^l$, which will be the adjustment subtracted from future output of the clock.
        This means that, if the clock offset were $\theta^l$ before synchronization, it is $\theta^l - \delta\theta^l$ after synchronization.
        We now derive safety bounds on $\delta\theta^l$ so that the clock is safe even with manipulated synchronization.

        We start with the safety condition {\em after synchronization} with Equation~\eqref{eq: safe sync cond}:
        \begin{equation} \label{eq: safe sync cond}
            -\frac \Theta 2 < \theta^l - \delta\theta^l < \frac \Theta 2.
        \end{equation}
        Equation~\eqref{eq: safe sync cond} enforces that the clock drift after synchronization meets the requirement of Equation~\eqref{eq: sync cond both} for loose time synchronization.
        With an NTS query, we can bound $\theta$ via Equation~\eqref{eq: theta bounds}.
        For convenience, we subtract $\delta\theta^l$ from all sides of Equation~\eqref{eq: theta bounds} to arrive at Equation~\eqref{eq: theta bounds sub star}:
        \begin{equation} \label{eq: theta bounds sub star}
            -(t_2-\tau_1) - \delta\theta^l < \theta^l - \delta\theta^l < \tau_4 - t_3 - \delta\theta^l.
        \end{equation}
        To derive safety bounds on $\delta\theta^l$, we combine Equations~\eqref{eq: safe sync cond}~and~\eqref{eq: theta bounds sub star}.

        Equations~\eqref{eq: safe sync cond}~and~\eqref{eq: theta bounds sub star} each express bounds on $\theta - \delta\theta^l$.
        However, since Equation~\eqref{eq: safe sync cond} is the safety condition the synchronization must satisfy and Equation~\eqref{eq: theta bounds sub star} is a bound on a measurement, the domain of Equation~\eqref{eq: safe sync cond} must form a superset of the domain of Equation~\eqref{eq: theta bounds sub star}.
        For the upper side, we derive the following lower bound on $\delta\theta^l$.
        \begin{align}
            \tau_4^l - t_3^l - \delta\theta^l &< \frac \Theta 2 \nonumber \\
            \tau_4^l - t_3^l - \frac \Theta 2 &< \delta\theta^l \label{eq: theta star lower}
        \end{align}
        For the lower side, we derive the following upper bound on $\delta\theta^l$.
        \begin{align}
            -\frac \Theta 2 &< -(t_2^l-\tau_1^l) - \delta\theta^l \nonumber \\
            \delta\theta^l &< -(t_2^l-\tau_1^l)+\frac \Theta 2 \label{eq: theta star upper}
        \end{align}
        Combining Equations~\eqref{eq: theta star lower}~and~\eqref{eq: theta star upper}, we form the bound of Equation~\eqref{eq: theta star bounds}.
        \begin{equation} \label{eq: theta star bounds}
            \tau_4^l - t_3^l - \frac \Theta 2 < \delta\theta^l < -(t_2^l-\tau_1^l)+\frac \Theta 2
        \end{equation}
        Any $\delta\theta^l$ that satisfies Equation~\eqref{eq: theta star bounds} will generate a safe clock at the moment of synchronization.
        Then, Equation~\eqref{eq: safe false alerts}~and~\eqref{eq: safe security} determine how long the clock synchronization is safe over time.

        Solving the case where there is no $\delta\theta^l$ that satisfies Equation~\eqref{eq: theta star bounds} derives an alert condition as a function of the synchronization's round trip time.
        To do this, one can flip the direction of the inequality in Equation~\eqref{eq: theta star bounds}, as in the following:
        \begin{align}
            \tau_4^l - t_3^l - \frac \Theta 2 &> -(t_2^l-\tau_1^l)+\frac \Theta 2, \\
            \Theta &< \tau_4^l - t_3^l + t_2^l-\tau_1^l. \label{eq: sync fail cond}
        \end{align}

        \subsubsection{Proof of Receipt Safety}

            To show safety in the synchronization protocol, we will show that if an adversary delays messages in the protocol, the adversary cannot fool a receiver into synchronizing to an unsafe condition.
            Here, the adversary desires to fool a receiver to select a $\delta\theta^l$ so that after synchronization, the clock is outside loose time synchronization.
            This corresponds to the condition $\theta^l - \delta\theta^l < - \frac \Theta 2$.

            According to Equation~\eqref{eq: theta star bounds}, the receiver may select any $\delta\theta$ that is in between those bounds (noting that a sensible receiver might elect to select the mid point of the bound).
            Let's assume that the receiver would make the choice most favorable to the adversary.
            Since the adversary wants $\theta^l - \delta\theta^l$ to be as small as possible, with the negative sign, the relevant bound from Equation~\eqref{eq: theta star bounds} is the upper bound.
            That is, we assume that the receiver selects the largest available $\delta\theta$ so as to make $\theta^l - \delta\theta^l$ as small as possible to assist the adversary in meeting their win condition.
            Because the bound involving $\tau_1$ and $t_2$ is the relevant one, we can reuse the attack diagram from Figure~\ref{fig: cert attack}.
            Again, the adversary may introduce $\Delta_{1,2}\geq 0$ and have $\varepsilon \rightarrow 0$.

            Starting from the unsafe condition, we rearrange for convenience:
            \begin{align}
                \theta^l - \delta\theta^l &< -\frac \Theta 2, \\
                \theta^l + \frac \Theta 2 &< \delta\theta^l.
            \end{align}
            Now, we apply the upper bound from Equation~\eqref{eq: theta star bounds}:
            \begin{align}
                \theta^l + \frac \Theta 2 &< -(t_2^l-\tau_1^l)+\frac \Theta 2, \\
                \theta^l &< -(t_2^l-\tau_1^l), \nonumber \\
                \theta^l &< -t_2^l+t_1^l + \theta^l, \nonumber \\
                0 &< -\Delta_{1,2}, \nonumber \\
                \Delta_{1,2} &< 0.
            \end{align}

            This contradicts the initial requirement that the adversary introduce $\Delta_{1,2}\geq 0$.
            Therefore, the adversary cannot fool a receiver into synchronizing into an unsafe condition.

            Suppose an adversary were to induce a $\Delta_{3,4} \geq 0$ on the NTS query return.
            This manipulation would increase the lower bound on $\delta\theta$ in Equation~\eqref{eq: theta star lower}.
            But since $\delta\theta$ still adheres to the upper bound of Equation~\eqref{eq: theta star lower}, the lower bound is not the active constraint on $\delta\theta$ in the proof of the smallest obtainable $\theta$ after synchronization under adversarial manipulation.

\section{Addressing NTS Vulnerabilities with TESLA-enabled GNSS} \label{sec: NTS changes}

    Suppose a receiver initiates any of the procedures of Section~\ref{sec: safe check and sync} but receives no response from the server.
    If the receiver does not shut down, the receiver is vulnerable to attack.
    This brittle shutdown condition could represent a major inconvenience, and some safety-of-life contexts may prohibit such a shutdown.
    This vulnerability with TESLA-enabled GNSS results from (1) the broadcast-only context and (2) disclosing or leaking information about $\tau_1$ in the synchronization protocol.

    In Section~\ref{sec: vuln desciption}, we describe how vulnerabilities arise and concretely describe how to attack a receiver.
    In Section~\ref{sec: NTP exp}, we apply our attack on data provided by an NTP server to show its feasibility.
    In Section~\ref{sec: address vuln}, we describe how to address the vulnerability (but not with the same level of surety as the above proofs).

    The vulnerabilities and methods of this section apply to a receiver unwilling to shut down if an attempted time-synchronization fails.
    After attempting a synchronization, information is leaked about the state of the receiver's GIC.
    The adversary can use this information to attack unsuspecting, broken receivers.
    If a receiver is willing to wait for service to check its clock using the methods of Section~\ref{sec: safe check and sync}, then the receiver will not suffer these vulnerabilities.
    However, we provide this section because we expect some contexts may need to continue operating when synchronization is denied.

    \subsubsection{Synchronization Vulnerability} \label{sec: vuln desciption}

        Unless NTS is modified according to Section~\ref{sec: address vuln}, NTS lends to a vulnerability resulting from (1) the broadcast-only context and (2) disclosing $\tau_1$.
        First, the loose-time synchronization condition for TESLA security was modified to accommodate broadcast-only GNSS.
        In {\em normal} TESLA, $\Theta$ is selected based on synchronization pings with Equation~\eqref{eq: TESLA bootstrap orig}.
        If a man-in-the-middle increases $t_2$ or a clock spontaneously increases its lag, $\Theta$ increases to prevent the acceptance of forged messages until there is a denial of service.
        In the GNSS context, $\Theta$ is constant for the entire constellation.
        This changes the safety condition to the enforcement of Equation~\eqref{eq: safe security}.
        Second, the act of distributing $\tau_1$ reveals information about the state of the receiver's clock.
        An adversary can use knowledge of a receiver's broken clock to submit forged messages to them and block the unsafe clock's detection.

        The mathematical arguments herein impose an assumption on the clock offset over time via Equation~\eqref{eq: safe security} and bounding function $B(\cdot)$.
        However, clocks are stochastic instruments.
        For instance, radiation could spontaneously flip a register, causing the GIC to go into a broken condition that would violate receipt safety.
        Depending on the clock certification, this would certainly be a rare force majeure, but it still can happen.
        The principal vulnerability here is that the adversary can identify such broken receivers, and finding just one could cause serious havoc.

\begin{figure}
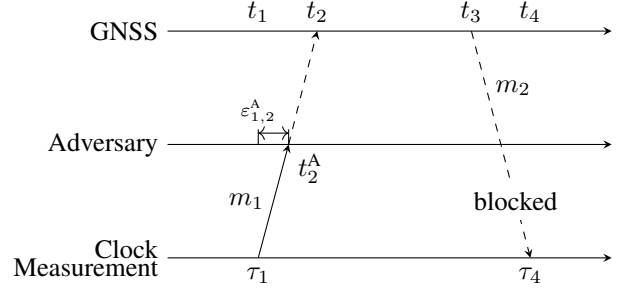

            \centering
            \tikzstyle{connect} = [inner sep=0pt, outer sep=0pt, circle, fill=black, minimum size=0pt]
            \def\lineLength{6}
            \def\eplength{0.6}
            \def\eventlength{1.2}   
            \tikz{
                \node[anchor=east] (GNSS) at (0, 0) {GNSS};
                \node[anchor=east, text width=2cm, align=right, font=\linespread{0.5}\selectfont] (AdvC) at (0, -1.5) {Adversary};
                \node[anchor=east, text width=2cm, align=right, font=\linespread{0.5}\selectfont] (RecC) at (0, -3) {Clock Measurement};

                \node (GNSSe) at (\lineLength, 0) {};
                \node (Adve) at (\lineLength, -1.5) {};
                \node (RecCe) at (\lineLength, -3) {};

                \draw[-stealth] (GNSS) -- (GNSSe);
                \draw[-stealth] (AdvC) -- (Adve);
                \draw[-stealth] (RecC) -- (RecCe);

                \node[connect] (t1) at (1*\eventlength, 0) {};
                \node[above] (lt1) at (t1) {$t_1$};

                \node[connect] (t2a) at (2*\eventlength - 0.35*1.9*\eventlength, -1.5) {};
                \node[below right] (lt2a) at (t2a) {$t_2^\text{A}$};

                \node[connect] (t2) at (2*\eventlength - 0.35*\eventlength, 0) {};
                \node[above] (lt2) at (t2) {$t_2$};

                \node[connect] (t3) at (3*\eventlength + 0.35*\eventlength, 0) {};
                \node[above] (lt3) at (t3) {$t_3$};

                \node[connect] (t4) at (4*\eventlength, 0) {};
                \node[above] (lt4) at (t4) {$t_4$};

                \node[connect] (tau1) at (1*\eventlength, -3) {};
                \node[below] (ltau1) at (tau1) {$\tau_1$};

                \node[connect] (tau4) at (4*\eventlength, -3) {};
                \node[below] (ltau4) at (tau4) {$\tau_4$};

                \draw[-stealth] (tau1) -- node[midway, left] {$m_1$} (t2a);
                \draw[-stealth, dashed] (t2a) -- (t2);
                \draw[-stealth, dashed] (t3) -- node[pos=0.25, right] {$m_2$} node[pos=0.75, fill=white] {blocked} (tau4);

                \draw[-] ([yshift=0.2cm]1*\eventlength, -1.5) -- ([yshift=-0cm]1*\eventlength, -1.5);
                \draw[-] ([yshift=0.2cm]2*\eventlength - 0.35*1.9*\eventlength, -1.5) -- ([yshift=-0cm]2*\eventlength - 0.35*1.9*\eventlength, -1.5);
                \draw[<->] ([yshift=0.15cm]1*\eventlength, -1.5) -- ([yshift=0.15cm]2*\eventlength - 0.35*1.9*\eventlength, -1.5);
                \node[anchor=south] at ([yshift=0.15cm]1*\eventlength, -1.5) {\scriptsize$\varepsilon_{1,2}^\text{A}$};
            }
            \caption{A conceptual diagram that depicts the structure of the attack on the vulnerability described in Section~\ref{sec: vuln desciption}. $t_1$ through $t_4$ and $\tau_1$ through $\tau_4$ have the same definitions from Figure~\ref{fig: nts}. $\varepsilon_{1,2}^\text{A}$ is the network transit time from the receiver to the adversary, and $t_2^A$ is the arrival time to the adversary of that message. As long as the return trip is blocked, the receiver will not be able to have any knowledge about the status of its clock.}
            \label{fig:attack}
        \end{figure}

        The attack is diagrammed with Figure~\ref{fig:attack}, described in Algorithm~\ref{alg: attack unencrypted}, and summarized as follows.
        An adversary forms a model on the receiver clock's $\theta$ and $\tau_1$ and listens to synchronization traffic.
        Eventually, a receiver clock will spontaneously drift into violating the loose time synchronization (when the $B(\cdot)$-clock model fails), and when the receiver attempts to synchronize, the traffic will be observed by the adversary.
        Using the model and information from the synchronization traffic, the adversary estimates with a high probability that a receiver is non-compliant and blocks the NTS traffic back to that receiver to stop the clock's broken condition detection and correction.
        The adversary knows that the receiver will not shut down with a denied NTS query return.
        The adversary knows with a high probability that that receiver will accept forged messages.

       \begin{algorithm}
            \caption{Attack on Unencrypted Traffic by Adversary}
            \label{alg: attack unencrypted}
            \begin{algorithmic}[1]
                \STATE Adversary begins observing the time-synchronization traffic of the vehicle class associated with a specific location to search for a vulnerable receiver.
                \STATE Adversary forms a model on the network traffic transit time from the receiver to the adversary (e.g., Equation~\eqref{eq: prob model}).
                \STATE Adversary eavesdrops on the NTS request $m_1 = (\eta, \tau_1, s^\textrm{receiver}_1)$ where $\tau_1$ is the time receiver recorded at the moment of sending $m_1$ and $s^\textrm{receiver}_1$ is an authentication signature on $(\eta, \tau_1)$.
                \STATE Adversary records $t_2^A$, the time of receipt of the eavesdropped message $m_1$.
                \STATE Using its internet traffic model, $t_1$ and $t_2^A$, Adversary observes traffic until it observes a receiver believed to have a $\theta < -\frac \Theta 2$.
                \STATE Adversary blocks the return NTS response $m_2=(\eta, \tau_1, t_2, t_3, s^\textrm{receiver}_2)$ and subsequent return NTS responses back to the receiver believed to accept forgeries so that that receiver's clock is never corrected.
                \STATE Adversary listens to the authentic GNSS signal for a disclosed TESLA key, generates a forged message and broadcasts it to the vulnerable receiver.
            \end{algorithmic}
        \end{algorithm}

        As an overview of the attack, Figure~\ref{fig:attack} diagrams the attack with the mathematical quantities for our description.
        We assume that the adversary and provider are perfectly synchronized to actual time, whereas the receiver may have an offset of $\theta$ that an adversary will hope is $\theta < -\frac \Theta 2$.
        The receiver will initiate an NTS query and note $\tau_1$ when it sends its initial request.
        The adversary will observe the NTS request and note its incoming time, which we denote as $t_2^A$.
        $\varepsilon_{1,2}^\text{A}$ is the signal travel time from the receiver to the intercepting adversary.
        The adversary can allow the NTS request to continue to the provider, but the adversary must block the return journey.
        If the NTS response is allowed back and then recorded at $\tau_4$, the receiver can correct its clock or know that the round trip time disqualifies using the query and initiates shutdown.

        To derive a model of the attack scenario, we start with  the events in the adversary frame with Equation~\eqref{eq: start}.
        Substituting the measurement equation, we arrive at Equation~\eqref{eq: model}:
        \begin{align}
            t_2^\text{A} &= t_1 + \varepsilon_{1,2}^\text{A}, \label{eq: start} \\
            t_2^\text{A} &= \tau_1 - \theta + \varepsilon_{1,2}^\text{A}, \nonumber \\
            \theta  &= - (t_2^A - \tau_1) + \varepsilon_{1,2}^\text{A}. \label{eq: model}
        \end{align}

        The adversary will use (a) evidence from the NTS traffic and (b) a model of the network traffic transit time to decide whether a specific receiver is vulnerable.
        In Equation~\eqref{eq: model}, $- (t_2^A - \tau_1)$ is (a), and $\varepsilon_{1,2}$ is (b).
        For (a), this is a search for outliers.
        We provide a simple estimation procedure of how long an adversary will need to search before finding a vulnerable receiver in Section~\ref{sec: Poisson Model}.

        For (b), a simple approach will suffice.
        Using a study of NTP network traffic, such as from \cite{practicalNTP}, we can come up with a correct but somewhat heuristic bound of Equation~\eqref{eq: heuristic bound}.
        We justify our bound since NTP round-trip-time is generally around 100 ms and expect the $\varepsilon_{1,2}$ to be around 50 ms, so we consider the model of Equation~\eqref{eq: heuristic bound} to be conservative.
        Nevertheless, an adversary can experientially adjust these considerations to the context:
        \begin{equation}\label{eq: heuristic bound}
            0.99 < \textrm{Prob}(\varepsilon_{1,2}^\text{A} < 100 \textrm{ms}).
        \end{equation}

        Substituting Equation~\eqref{eq: model} into Equation~\eqref{eq: heuristic bound}, we arrive at Equation~\eqref{eq: prob model}.
        \begin{equation}\label{eq: prob model}
            0.99 < \textrm{Prob}(\theta < - (t_2^\text{A} - \tau_1) + 100 \textrm{ms})
        \end{equation}

        Using Equation~\eqref{eq: prob model}, if an adversary observes traffic where $- (t_2^\text{A} - \tau_1) + 100 \textrm{ms}  < -\frac \Theta 2$, then that particular receiver is the broken state of $\theta < - \frac \Theta 2$ with high probability.

        $\tau_1$ would not be available in the clear if the NTS traffic were encrypted and the server graciously agreed never to leak or disclose $\tau_1$ to malicious parties.
        Or if the receiver simply omitted $\tau_1$ from the protocol.
        When $\tau_1$ is not in the clear, the attack model can be modified to incorporate receiver implementation procedures that leak information about $\tau_1$.
        This would involve replacing $\tau_1$ with $\text{Prob}(\tau_1 \mid t_2^A)$.
        For instance, if a receiver was known to initiate NTS requests at a specific interval (e.g., the top of the minute), then one could form a model of $\text{Prob}(\tau_1 \mid t_2^A)$ that involves judiciousy mapping (e.g., rounding to the nearest minute) $t_2^A$ to likely $\tau_1$.
        It would be apt for the receiver to introduce randomness to force the adversary to form a {\em uniform} model $\text{Prob}(\tau_1 \mid t_2^A)$ to limit the information gleaned from $t_2^A$.

        Solutions that claim to minimize $\tau_1$ information leakage should consider the following scenario.
        First, the adversary can observe the presence of a successful NTS synchronization and infer that the clock is synchronized to $\theta=0$ at that time.
        The adversary has all relevant clock technical specifications, meaning it can compute the next time of synchronization $t$ using the clock's specification of $B(\cdot)$.
        The adversary could infer that the next $\tau_1$ is the computed next synchronization time $t$ from the GIC specification of $B(\cdot)$.
        In other words, after observing the last successful synchronization, the adversary can assume that the next synchronization will result from a $\tau_1$ equal to the time derived from its clock drift bound function or some other non-random specification.
        Moreover, the adversary can block synchronization traffic, forcing the receiver to take worst-case procedures.
        For instance, an adversary could refuse service to a receiver until the time approaches that last acceptable synchronization time to generate the best possible $\text{Prob}(\tau_1 \mid t_2^A)$.
        In aggregate, one should assume that the adversary knew the last time when $\theta=0$ and knows the nominal synchronization time of $\tau_1$.

        We should be concerned about the situation where the adversary cares about spoofing {\em any} receiver rather than a particular receiver.
        In the context of the large vehicles potentially employing TESLA-enabled GNSS (e.g., autonomous cars and planes), finding a single vulnerable vehicle would be enough to wreak havoc.
        One could assert that, based on the assumption that the clock drift is bounded by $B(\cdot)$, the receiver should know to shut itself off before the clock could ever violate loose-time synchronization.
        However, we consider the following two situations.
        First, given the context of flight, force-majure events will occur, such as radiation spontaneously changing a clock time.
        Second, the vehicle context will frustrate shut-down requirements (e.g., we expect resistance to conservative requirements causing flight or ride-share delay resulting from a busy time synchronization server).
        Therefore, it may be best to implement the mitigations presented in Section~\ref{sec: address vuln}.

    \subsubsection{Experimental Observation} \label{sec: NTP exp}

        Via private correspondence, the authors of \cite{sherman2016usage} graciously gave us access to the $\tau_1$ and $t_2$ values from a 2015 study regarding NTP server usage.
        The data provided does not contain identifiable information among the NTP users (just the time stamps in the NTP messages).
        Using that data and a conservative NTP round trip time model, we found large numbers of users that likely had TESLA-non-compliant clocks were they using a future TESLA-enabled GNSS~\cite{AndersonnJournal}.
        We emphasize the careful interpretation of the meaning of these results, limiting our conclusion to just that our attack is immediately feasible.

        The population of timestamps reflects the varying inconsistencies of users not faithfully implementing the NTP protocol.
        Among the approximate 12.7 million requests analyzed, about 26\% transmitted a null $\tau_1$ (i.e., $\tau_1 = 0$), and about 5\% transmitted an course integer second $\tau_1$.
        The mode of the $\tau_1$ centered about the correct time, indicating the majority of users are leaking the lagging state of their clock.
        While the data was collected in 2015, there were a large number of $\tau_1$ with the calendar year of 1970 (likely from the unix epoch).
        Moreover, a substantial section of users form an apparent uniform floor.
        We suspect this uniform band results from users transmitting a random $\tau_1$, perhaps as a nonce to differentiate repeated requests.
        Lastly, we note that we observe elevated bursts, such as the top of the hour and top of the minute, which could reveal information about specific requests' $\tau_1$.

\begin{figure}
            \centering
            \includegraphics[width=\linewidth]{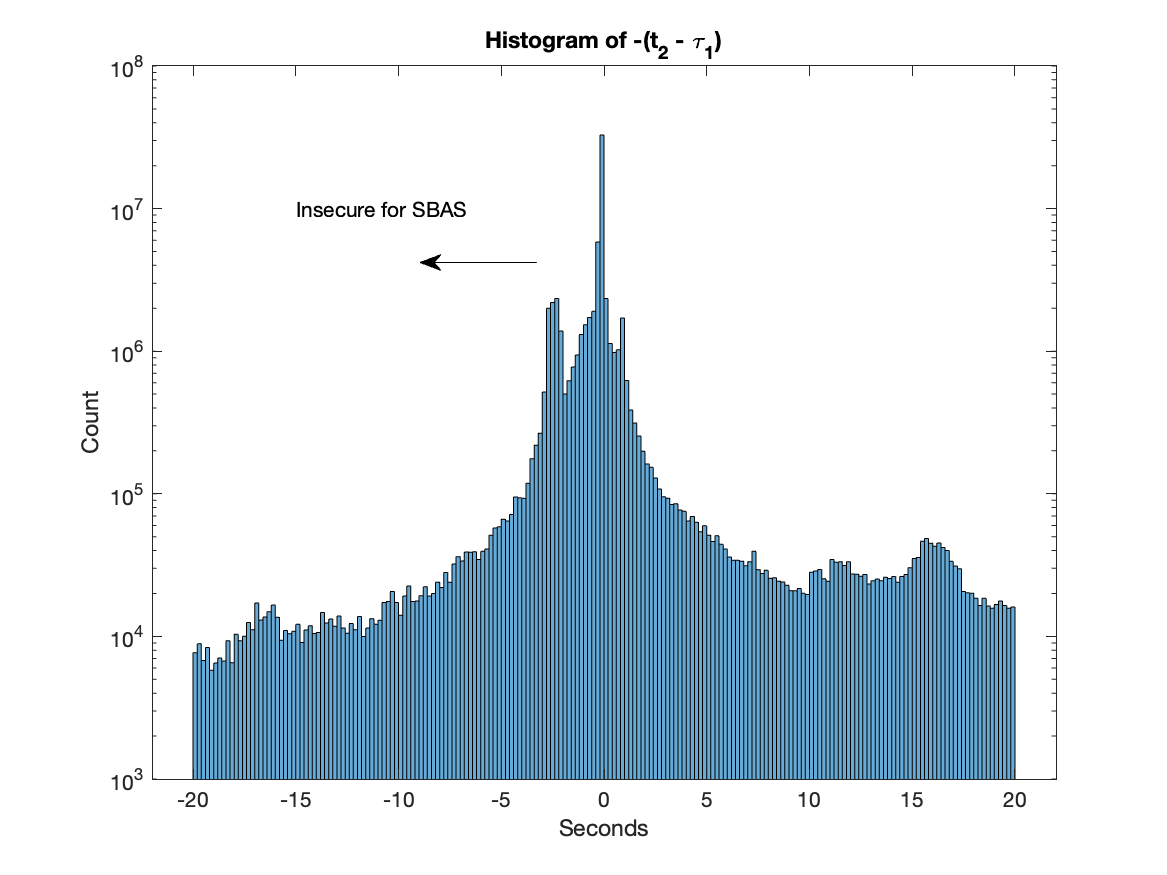}
            \caption{
                The distribution of $-(t_2-\tau_1)$ zoomed into the mode of the distribution around near-synchronized clocks.
                Neglecting a conservative 100ms NTP request transmission time, a negative value indicates that a particular clock is lagging.
                If the user were to use a TESLA-enabled GNSS, a value less than $-\frac \Theta 2$ indicates that the user would accept forgeries.
                We annotate where future TESLA-enabled SBAS users of ~\cite{AndersonnJournal} would accept forgeries.
            }
            \label{fig: close up histogram}
        \end{figure}

        We provide a close-up of the distribution of $-(t_2-\tau_1)$ values around the correct time in Figure~\ref{fig: close up histogram}.
        We find a substantial number of users who, ostensibly, are providing a faithful $\tau_1$ value, and we observe a substantial incidence of users whose clocks are likely lagging behind.
        In Figure~\ref{fig: close up histogram}, we annotate which users, if they were listening to a future authenticated SBAS concept with $\Theta=6$~\cite{AndersonnJournal}, would accept forgeries.

        From these results, we make the following {\em limited} claim.
        Since we, the authors, using data provided by the back-end of an NTP server, can observe clocks of users who would, with high probability, accept forgeries in a TESLA-enabled GNSS were they using that TESLA-enabled GNSS, the NTP server would certainly be able to identify these vulnerable users.
        Moreover, given our threat model and how the traffic is unencrypted, an adversary would also be able to identify vulnerable receivers immediately.
        The adversary's game is to wait for a vulnerable user.
        We show the concrete plausibility of exploiting the vulnerability we point out.
        Therefore, this vulnerability should concern those wishing to provide authentication security via a TESLA-enabled GNSS.

        We must limit the interpretation beyond a demonstration.
        Interpretation of the actual data should be limited given the wide variety of actual uses of $\tau_1$.
        Moreover, we do not warrant that this traffic is representative of future synchronization traffic for a future TESLA-enabled GNSS.
        We would have no indication whether the time $\tau_1$ is measured from the clock used to enforce the TESLA-loose-time synchronization requirement or a separate clock.
        Furthermore, we do not expect the drift rates of receiver clocks to be consistent with the drift rate of clocks from this NTP traffic.


        \subsubsection{Poisson Model} \label{sec: Poisson Model}

            An adversary must wait to find a vulnerable clock.
            Depending on the receiver clock drift rate, the incidence might be similar to that of rare outliers.
            However, by characterizing the data, an adversary can reasonably model how long they will expect to wait, such as with a Poisson model.

            Suppose that this data came from users using a TESLA-enabled SBAS with $\Theta=6$, as in \cite{AndersonnJournal}.
            Considering the wide variety of $\tau_1$ observed, we assume incoming NTP requests received within 20 seconds of real-time contain $\tau_1$ that reflects their real clock.
            Therefore, among those requests, requests with a $-(t_2-\tau_1) < -3.1$ should accept forgeries with high probability.
            Taking the $t_2$ from those offending requests, we form a maximum-likelihood-estimated Poisson model on the incidence of unsafe receivers using the NTP service.
            We calculate a Poisson model with $\lambda = 0.57$ seconds.
            Therefore, on expectation, for this data, the adversary should expect to encounter a vulnerable clock after 0.57 seconds of observation.
            There were about 7000 requests per second in this data set, which handily suggests that vulnerable clocks are certainly outliers. 
            Moreover, we also note that we expect vulnerable clocks among those who did not send a faithful $\tau_1$. 

    \subsubsection{Addressing NTS Vulnerabilities} \label{sec: address vuln}

        To mitigate the threat posed by Section~\ref{sec: vuln desciption}, a receiver must limit information leakage of $\tau_1$.
        This can be done with two simple modifications to NTS.
        First, $\tau_1$ should be omitted from NTS messages or replaced with a nonce.
        Second, NTS queries should occur uniformly randomly over a large enough interval.

        Regarding the first modification, we already observed users doing this in Section~\ref{sec: NTP exp}.
        Removing $\tau_1$ from the protocol poses no burden to the synchronization because the non-receiver party does not need $\tau_1$ in any computation.
        If one were to encrypt the NTS traffic, an eavesdropper would not be able to engage in this attack; however, the synchronization server could leak $\tau_1$ or use it as an adversary itself.
        In Section~\ref{sec: NTP exp}, the server did this so we could demonstrate this attack.
        Therefore, it would be best to not trust the server and simply omit $\tau_1$ or replace $\tau_1$ with an unrelated value.

        Regarding the second modification, NTS queries that occur on a predictable basis leak information about $\tau_1$.
        A good counter-action to this security concern is to ensure that the best model an adversary can form on $\text{Prob}(\tau_1 \mid t_2^A)$ is a uniform distribution with support exceeding anything useful to the adversary.
        A uniform $\text{Prob}(\tau_1 \mid t_2^A)$ translates to a uniform adversarial-deduced distribution of $\theta$ given the evidence from adversarial eavesdropping.

\begin{figure}
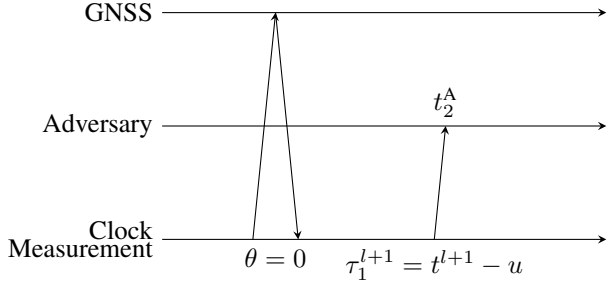

            \centering
            \tikzstyle{connect} = [inner sep=0pt, outer sep=0pt, circle, fill=black, minimum size=0pt]
            \def\lineLength{6}
            \def\eplength{0.6}
            \def\eventlength{1.2}   
            \tikz{
                \node[anchor=east] (GNSS) at (0, 0) {GNSS};
                \node[anchor=east, text width=2cm, align=right, font=\linespread{0.5}\selectfont] (AdvC) at (0, -1.5) {Adversary};
                \node[anchor=east, text width=2cm, align=right, font=\linespread{0.5}\selectfont] (RecC) at (0, -3) {Clock Measurement};

                \node (GNSSe) at (\lineLength, 0) {};
                \node (Adve) at (\lineLength, -1.5) {};
                \node (RecCe) at (\lineLength, -3) {};

                \draw[-stealth] (GNSS) -- (GNSSe);
                \draw[-stealth] (AdvC) -- (Adve);
                \draw[-stealth] (RecC) -- (RecCe);

                \node[connect] (t1) at (1.25*\eventlength, 0) {};
                \node[below, yshift=-3cm] (lt1) at (t1) {$\theta=0$};

                \node[connect] (tau1) at (1*\eventlength, -3) {};
                \node[connect] (taua) at (1.5*\eventlength, -3) {};

                \node[connect] (tau4) at (3*\eventlength, -3) {};
                \node[below] (ltau4) at (tau4) {$\tau_1^{l+1} = t^{l+1} - u$};

                \node[connect] (t2a) at (3.125*\eventlength, -1.5) {};
                \node[above] (lt2a) at (t2a) {$t_2^\text{A}$};

                \draw[-stealth] (tau1) -- (t1);
                \draw[-stealth] (t1) -- (taua);
                \draw[-stealth] (tau4) -- (t2a);
            }
            \caption{
                A conceptual diagram of the situation where an adversary uses the time of the last synchronization to predict $\tau_1$ to form a model of Prob($\theta \mid t, t_2)$.
                To mitigate this attack and limit the information leakage from synchronization itself, we suggest the receiver introduce uniform randomness into the NTS query initiation.
                This ensures that an adversarial model of Prob($\theta \mid t, t_2)$ must be uniform.
                The transmission lines approach vertical to reflect the case when the adversary is immediately adjacent to the receiver to achieve a conservative case and simplify math.
            }
            \label{fig: attack last}
        \end{figure}

        As in the conceptual diagram of Figure~\ref{fig: attack last}, suppose that the last synchronization occurred at $t^l$ and, using Equations~\eqref{eq: safe false alerts}~and~\eqref{eq: safe security}, the receiver computes the next $t$ for synchronization.
        Suppose the receiver initiates the next synchronization at $\tau_1^{l+1} = t - u$ with $u \leftarrow U(0, 2\lambda\Theta)$.
        $\lambda \geq 1$ is a slack parameter that indicates how much support we will allow with the adversarial model.
        $\lambda = 1$ corresponds to the case where the spread is equal to the allowed drift assuming $B(\cdot)$ in between synchronizations.

        We show that, given the adversary-observed evidence, the best model on $\theta$ is uniform.
        Suppose that the adversary is immediately adjacent to the receiver, meaning the transit time from receiver to adversary is instantaneous, forming a conservative model while simplifying math.
        With Equation~\eqref{eq: instant last attack}, we derive Equation~\eqref{eq: theta uniform model} by substituting the receiver's selected resynchronization time from the proceeding paragraph.
        \begin{align}
            t_1^{l+1} &= t_2^{l+1} \label{eq: instant last attack} \\
            \tau_1^{l+1} - \theta^{l+1} &= t_2^{l+1} \nonumber \\
            t - u -\theta^{l+1} &= t_2^{l+1} \nonumber \\
            \theta^{l+1} &= t - t_2^{l+1} - u \label{eq: theta uniform model}
        \end{align}
        From Equation~\eqref{eq: theta uniform model}, $t-t_2$ is known to the adversary.
        Since $u$ is uniform and unknown to the adversary, the distribution Prob($\theta^{l+1} \mid t-t_2^{l+1})$ is uniform over $(t-t_2^{l+1}, t-t_2^{l+1}-2\lambda\Theta)$.

        Our suggestion suffers from two issues.
        First, it's a necessity after several possibly assumed-true conditions are false.
        These include (a) that the clock spontaneously drifted outside the already assumed bound $B(\cdot)$ and (b) the receiver continued to operate beyond its resynchronization specification after being denied service by NTS.
        However, we note that this entire discussion is about the identification of outlying and adverse events for contexts that may not have the option to shut down (e.g., moving or flying vehicles).
        Second, our suggestion does not provably assure no information leakage: consider the case when the next NTS ping comes so egregiously late, ensuring that the entire spread of uniform Prob($\theta^{l+1} \mid t-t_2^{l+1})$ is in a broken state.
        As the clock approaches the boundary $\theta < -\frac \Theta 2$ and the support of uniform Prob($\theta^{l+1} \mid t-t_2^{l+1})$ includes some broken-clock domain, an adversary can compute the probability that a receiver is broken given the evidence.

        A receiver could choose a $\lambda = \frac{t}{2\Theta}$ so that a receiver is always uniformly querying their clock, producing a Prob($\theta^{l+1} \mid t-t_2^{l+1})$ with the maximal support confusing the adversary.
        However, an adversary could block all NTS attempts until the clock nears its specification time $t$, undoing a large-uniform-support strategy.
        We suspect there is no provable way to initiate an NTS query that admits no information leakage on $\tau_1$ with a single GIC.
        To provide provable safety against this attack, a receiver must shutdown if an NTS query ever fails.

\section{Experimental Validation}

    In this section, we validate the checks proposed in Sections~\ref{sec: clock check}~and~\ref{sec: safe check and sync} by simulation.
    To simulate, we generate a representative set of scenarios among $\theta$ and $\Delta$.
    In our simulations, we set $\Theta = 1$, meaning the forgery condition is $\Delta \geq 1$.
    We simulate clocks with offsets between -2 and 2.
    In addition to the adversary-induced delays, we add a 0.01 transmission latency within every simulated transmission step (e.g., to/from the receiver, adversary, and server, and during the server and adversary computational processing).

\begin{figure}
        \centering
        \includegraphics[width=\linewidth]{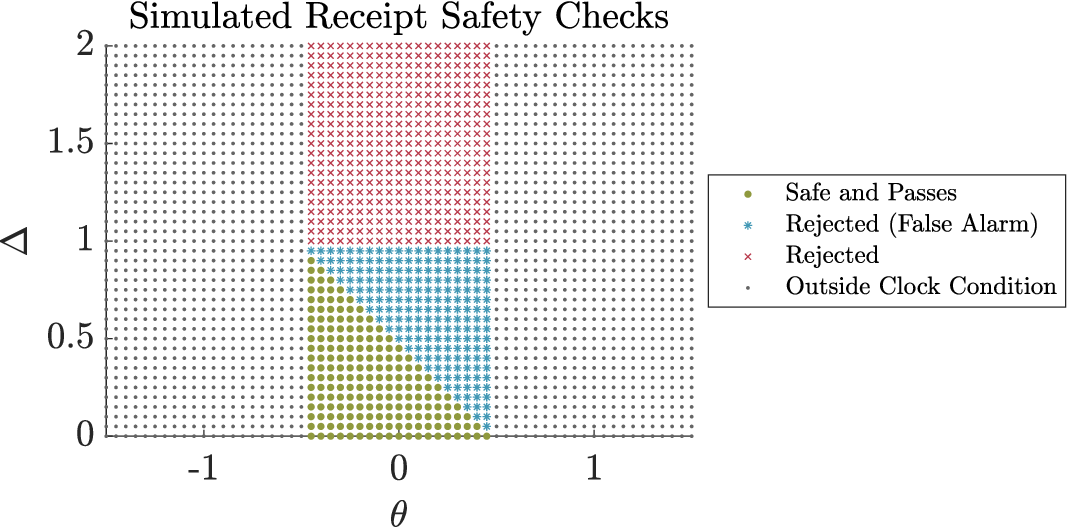}
        \caption{Results from simulations using a representative set of $\theta$ and $\Delta$ to validate the receipt safety check of Equation~\eqref{eq: th assert condition}.}
        \label{fig: receipt safety check}
    \end{figure}

    In Figure~\ref{fig: receipt safety check}, for each $\theta$ and $\Delta$, we evaluate the check of Equation~\eqref{eq: th assert condition}.
    Within the bounds of an acceptable clock synchronization, we observe no condition lending to the acceptance of a forged message.

    For the proofs related to verifying clock safety and ensuring safe synchronization, the active constraints involve $\tau_1, t_2$ and $\Delta_{1,2}$.
    This corresponds to proving conditions in the worst case $\theta$ (favorable to the adversary).
    However, a real spoofer will want to induce a $\Delta_{3,4}$.
    With $\Delta_{3,4}$, the spoofed measurement will appear to the receiver as if its clock is more {\em ahead} than reality.
    The spoofer wants a broken lagging clock to believe it is further ahead than reality so that it can fool the receiver of a lagging clock undetected.
    This means that the adversary inducing a $\Delta_{1,2} \geq 0$ will aid the receiver in detecting its unsafe condition and inducing a $\Delta_{3,4} \geq 0$ will help the adversary remain undetected when a clock starts lagging to a broken state.

\begin{figure}
        \centering
        \includegraphics[width=\linewidth]{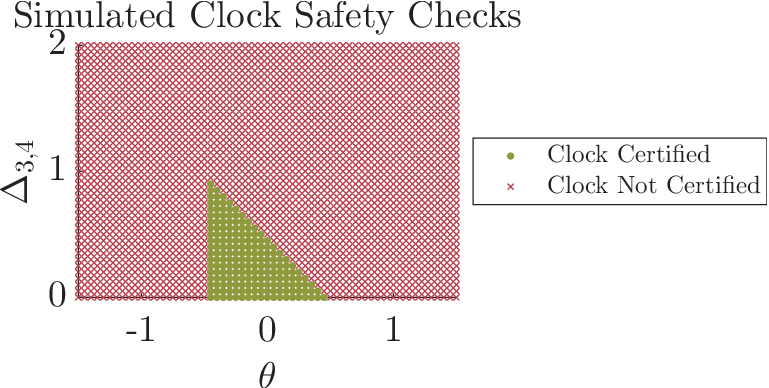}
        \caption{Results from simulations used to validate the clock safety check of Equations~\eqref{eq: safe false alerts}~and~\eqref{eq: safe security}. The simulation was conducted using a representative set of $\theta$ and $\Delta$.}
        \label{fig: clock safety check}
    \end{figure}

    In Figure~\ref{fig: clock safety check}, for each $\theta$ and $\Delta_{3,4}$, we evaluate the check of Equations~\eqref{eq: safe false alerts}~and~\eqref{eq: safe security} by simulation.
    Under no circumstances is a clock outside the synchronization conditions or in forgery conditions accepted with this check.
    We note that as the clock offset approaches the lagging unsafe condition, the amount of delay allowed to the adversary before detection increases.

\begin{figure}
        \centering
        \includegraphics[width=\linewidth]{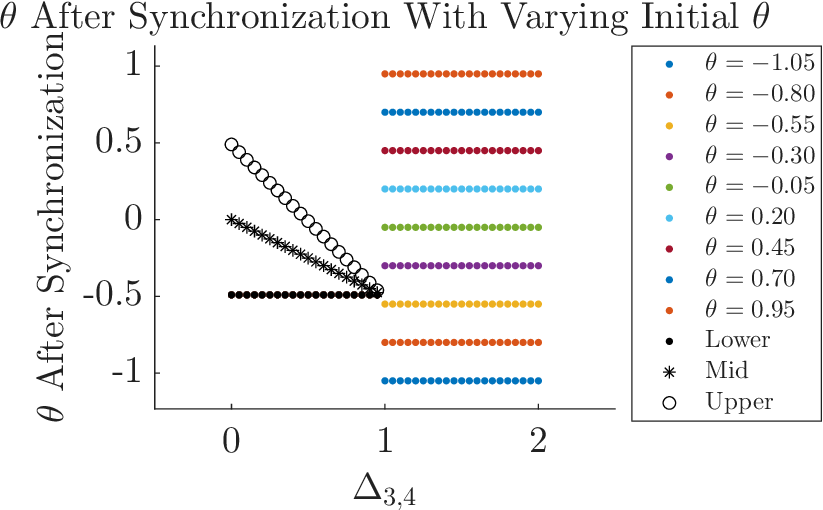}
        \caption{
            Results from simulations used to validate the synchronization safety of Equation~\eqref{eq: theta star bounds}. 
            The simulation was conducted using a representative set of $\theta$ and $\Delta$.
            The markers are black where all the scenarios converge.
            The dots presume the receiver elects the upper bound of Equation~\eqref{eq: theta star bounds} to help the adversary win the game.
            A sensible receiver will elect to select the midpoint of Equation~\eqref{eq: theta star bounds}, which are the starred markers.
            When $\Delta \geq 1$, the adversary may induce forgeries, but the receiver knows to not adjust the clock offset and shutdown; hence, the markers on the right part of the diagram reflect the initially simulated $\theta$.
        }
        \label{fig: after sync check}
    \end{figure}

    In Figure~\ref{fig: after sync check}, for each $\theta$ and $\Delta_{3,4}$, we evaluate $\theta$ {\em after} synchronization according to both bounds and midpoint of Equation~\eqref{eq: theta star bounds}.
    The receiver's selection of the upper bound is most favorable to the adversary resulting in the lowest $\theta$ after synchronization.
    In the figure, when $\Delta > \Theta$, $\tau_4 - t_3 + t_2-\tau_1 > \Theta$, where the receiver knows not to make any changes to the clock and stop authenticating.
    Hence, for $\Delta > 1$, the simulated clock offsets are not changed, reflecting how they were initially set in our simulations.
    For $\Delta < 1$, the clocks are set to the same state, which is why they are marked black in that region.
    The dots correspond to the worst case with the upper bound of Equation~\eqref{eq: theta star bounds}.
    The stars correspond to the sensible midpoint of the bounds of Equation~\eqref{eq: theta star bounds}, which is Equation~\eqref{eq: nts}.
    And the circles correspond to the lower bound of Equation~\eqref{eq: theta star bounds}.
    We show each to show that the adversary's manipulation pushes $\theta$ back, which will make the receiver more likely to not detect its lagging condition, but not enough to allow a forgery.

    In all cases, we observe what the proofs already show: that these methods will ensure that a receiver correctly ensure receipt safety.

\section{Addressing Multi-cadence TESLA Time Synchronization}

    There are GNSS authentication proposals that will include multiple, independent authentication schemes~\cite{Anderson2017}.
    The schemes will operate at different times to authentication (i.e., different $\Theta$) to accommodate different receiver use cases.
    For instance, there are two authentication TESLA instances for users with and without an internet connection~\cite{chimeraicd}.
    These two use cases indirectly determine the appropriate precision for the GIC: a receiver without an internet connection might need a low-drift clock.
    Or, a signal scheme could accommodate standard drift rates and lower frequency synchronizations by increasing the times to authenticate $\Theta$.
    A clock becomes vulnerable after its drift accumulates into lagging beyond provider time by $\frac \Theta 2$; therefore, increasing $\Theta$ decreases that likelihood and increases the time between maintenance.

    This section addresses whether a slower TESLA instance can assist or bootstrap a faster TESLA instance.
    It cannot.
    Suppose that the slower TESLA instance has $\Theta_\text{blue}$ and the faster has $\Theta_\text{red}$ (i.e., $\Theta_\text{red} < \Theta_\text{blue}$).
    We show that a receiver whose clock is rated for the slower TESLA can never assert the security of the faster TESLA instance.
    A slow TESLA scheme could redundantly sign the fast TESLA instance to save bandwidth for slow-TESLA receivers~\cite{anderson2022efficient}.
    In that case, there is no time-to-authentication advantage to the slow-TESLA receiver.
    In other words, meeting the loose-time synchronization requirement for slow TESLA and successfully authenticating a slow-TESLA message does not aid in satisfying the fast-TESLA loose-time synchronization requirement to utilize faster authentication.
    A receiver rated for a slow-TESLA instance is forever required to wait for the slow-TESLA delay release time asserting receipt safety.

\begin{figure}
        \includegraphics[width=\linewidth]{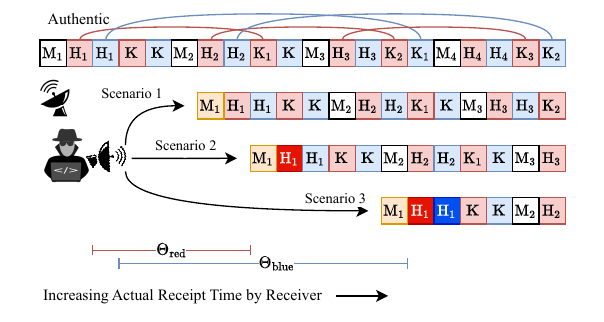}
        \caption{
            An adversary listens to the authentic signal and then replays a spoofed signal under three scenarios described in Table~\ref{tab: scenarios}.
            The signal contains two independent TESLA instances: red and blue.
            Each TESLA signal signs each message.
            The indices and the lines in the Authentic scenario connect an HMAC commitment to the corresponding delay-released key.
            The diagram reads from left to right in the order of increasing time of receipt by the receiver.
            For instance, a receiver in Scenario 2 receives the signal with greater delay than in Scenario 1.
            If the delay is large enough, then an adversary can listen for a key and forge an HMAC, which is highlighted in Scenarios 2 and 3.
            Table~\ref{tab: scenarios} describes the state of the GIC when the delay is not detected.
        }
        \label{fig: scenarios}
    \end{figure}

    To demonstrate the no-time-to-authenticate-advantage property of multiple TESLA instances, we create an illustrative scenario with two TESLA instances in the same signal.
    We provide a conceptual diagram with Figure~\ref{fig: scenarios} with accompanying Table~\ref{tab: scenarios}.
    In our illustrative scenario, a signal contains two TESLA instances: red and blue.
    Each message is signed by both TESLA instances.
    The key release for the red TESLA instance comes one m-h-k tuple later, and the key release for the blue TESLA instance comes two m-h-k tuples later.
    In our notation, $\Theta_\text{red} < \Theta_\text{blue}$.

    \begin{table}
        \centering
        \begin{tabular}{|C{30pt}|C{73pt}|p{115pt}|}
        \hline
        \textbf{Scenario} & \textbf{Broken Clock Condition} & \textbf{Attack Description} \\
        \hline
            1 & $-\frac {\Theta_\text{blue}}{2} < -\frac{\Theta_\text{red}}{2} < \theta$ & The adversary delays the signal by less than $\Theta_\text{red}$. Any forgery of either the red or blue $H_1$ will be detected. \\
        \hline
            2 & $-\frac {\Theta_\text{blue}}{2} < \theta < -\frac {\Theta_\text{red}}{2}$ & The adversary delays the signal greater than $\Theta_\text{red}$ and less than $\Theta_\text{blue}$. The adversary listens for the red $K_1$ and computes a forged red $H_1$ to sign a spoofed $M_1$. Any forgery of the blue $H_1$ will be detected upon release of the blue $K_1$. \\
        \hline
            3 & $\theta < -\frac{\Theta_\text{blue}}{2} < -\frac{\Theta_\text{red}}{2}$ & The adversary delays the signal greater than $\Theta_\text{blue}$. The adversary listens for the red and blue $K_1$ and computes a forged red and blue $H_1$ to sign a spoofed $M_1$. The spoof is never detected. \\
        \hline

        \hline
        \end{tabular}
        \caption{A description of the scenarios from Figure~\ref{fig: scenarios}. If a clock's condition is broken, then an attacker can engage in an attack that breaks authentication security in the corresponding scenario.}
        \label{tab: scenarios}
    \end{table}

    Scenario 2 presents the security flaw by incorrectly believing that a successful slow-TESLA authentication allows a slow-TESLA receiver to authenticate at the cadence afforded to fast-TESLA receivers.
    A receiver in a safe condition for a slow-TESLA instance but in a broken condition for the fast-TESLA instance would accept a fast-TESLA forgery.
    We expect that in real TESLA-enabled GNSS, slow-TESLA instances will redundantly sign fast-TESLA instances~\cite{anderson2022efficient} to save bandwidth.
    A slow-TESLA receiver checking the fast-commitments {\em and} the slow-commitments would detect the forgery in Scenario 2 at the slow-TESLA pace.
    Hence, we say that there is no time-to-authentication advantage, and a receiver should withhold its authenticated flag until then.
    If the slow-TESLA receiver in Scenario 2 does not check the slow-TESLA commitments, then it would accept forgeries.
    We provide proof of the existence of such an attack in Appendix~\ref{app: proof} to compute the $\theta$ and $\Delta$.

\section{Receiver Implementation Algorithms}

    For the implementor's convenience, we concretely aggregate all the concepts described in this work into Algorithms~\ref{alg: correct time sync}~and~\ref{alg: mhk receipt}.

    Algorithm~\ref{alg: correct time sync} contains the following features.
    First, a modified NTS query that accounts for the vulnerabilities resulting from disclosing $\tau_1$ (see Section~\ref{sec: NTS changes}).
    Second, the clock drift correction is bounded according to Section~\ref{sec: sync}.
    Third, the next synchronization time is computed to always satisfy safety conditions between synchronizations according to Section~\ref{sec: cert}.
    We refer to Section~\ref{sec: NTS changes} for a suggestion on how to modify this procedure if it is not acceptable to shutdown after a denial of synchronization service.

    \begin{algorithm}
       \caption{GIC Synchronize}
        \label{alg: correct time sync}
        \begin{algorithmic}[1]
            \STATE GNSS provider and receiver establish an asymmetrically encrypted and authenticated channel for NTS synchronization $l$.
            \STATE Receiver draws a nonce $\eta^l$ to associate the return message and deter replay and denial of service.
            \STATE Receiver sends message $m_1^l = (\eta^l, s^{\textrm{receiver}_1, l})$ {\em specifically omitting} $\tau_1$ or replacing the field with any value independent of $\tau_1^l$, and $s^{\textrm{receiver}, l}_1$ as the receiver's authentication signature on $(\eta^l,)$.
            \STATE Receiver measures the moment of sending message $m_1^l$ as $\tau_1^l$ and holds $\tau_1^l$ in {\em strict confidence}.
            \STATE Provider records $t_2^l$, the time of receipt of the message $m_1^l$.
            \STATE Provider sends message $m_2^l=(\eta^l, t_2^l, t_3^l, s^{\textrm{provider},l}_2)$ back to receiver, where $s^{\textrm{provider},l}_2$ is provider's authentication signature on $(k, t_2^l, t_3^l)$.
            \STATE Receiver records $\tau_4^l$ at the moment of receipt of message $m_2^l$.
            \IF{$\Theta < \tau_4^l - t_3^l + t_2^l-\tau_1^l$ from Equation~\eqref{eq: sync fail cond}}
                \STATE Receiver cannot authenticate any future messages and has leaked information about its clock offset (noting the alternative presented in Section~\ref{sec: NTS changes} and its limitations on provable safety).
                \STATE \textbf{return}
            \ENDIF
            \STATE The receiver adjusts its GIC by subtracting any $\delta\theta^l$ from its current GIC output that satisfies 
            \begin{equation} \tag{\ref{eq: theta star bounds}}
                \tau_4^l - t_3^l - \frac \Theta 2 < \delta\theta^l < -(t_2^l-\tau_1^l)+\frac \Theta 2.
            \end{equation}
            We suggest the of midpoint of the bounds in Equation~\eqref{eq: theta star bounds}: $\delta\theta = \frac 1 2 \cdot (\tau_4^l - t_3^l - t_2^l + \tau_1^l)$.
            \STATE The receiver computes the latest acceptable next synchronization time $t$ by solving the following program using constraint Equations~\eqref{eq: safe false alerts}~and~\eqref{eq: safe security} adjusted by the $\delta\theta^l$ correction of the previous step.
            \begin{align*}
                t^{l+1} = \quad & \max \quad t \\
                \textrm{subject to} \quad &\tau_4^l - \delta\theta^l - t_3^l + B(t - t^l) < \frac \Theta 2 \\
                & - \frac \Theta 2 < -(t_2^l - \tau_1^l + \delta\theta^l) - B(t - t^l)
            \end{align*}
        \end{algorithmic}      
    \end{algorithm}

    Algorithm~\ref{alg: mhk receipt} contains the procedures to ensure the receipt safety of the messages within broadcast-only TESLA correctly.
    If a message has receipt safety, the receiver must also perform the additional TESLA checks to ensure message integrity and authenticity.

    \begin{algorithm}
        \caption{Message Authentication within Broadcast-only TESLA}
        \label{alg: mhk receipt}
        \begin{algorithmic}[1]
            \STATE Receiver presumes its GIC satisfies the broadcast-only TESLA loose-time synchronization condition of Equation~\eqref{eq: sync cond both} via previous successful execution of Algorithm~\ref{alg: correct time sync}.
            \STATE Receiver uses the system design to correctly associate messages, commitments, and delay-released keys as (message $m$, commitment $h$, key k) Tuples.
            \FORALL{(message m, commitment h, key k) Tuples}
                \STATE Measure the receipt time of commitment h as $\tau_\text{h}$ and message m as $\tau_\text{m}$ with the GIC.
                \STATE Recall the correct release time of the corresponding key k as $t_\text{k}$.
                \IF{$\max (\tau_\text{h}, \tau_\text{m}) < t_\text{k} - B_l$}
                    \STATE m has receipt safety. Perform the additional TESLA security checks (e.g., h $=$ H(k, m)) to determine message integrity and message authenticity.
                \ELSE
                    \STATE m does not have receipt safety, so m does not have message integrity or authenticity.
                \ENDIF
            \ENDFOR
        \end{algorithmic}
    \end{algorithm}

    Algorithms~\ref{alg: correct time sync}~and~\ref{alg: mhk receipt} contain the two algorithms required from a timing perspective.
    As discussed in Section~\ref{sec: TESLA sync intro} and better-described in~\cite{AndersonnJournal}, the receiver must also check that the incoming key k hashes down to a hash signed with an asymmetric authentication instance that matches the provider.

\section{Conclusion}

    In this work, we discuss the timing algorithms of the GIC required to assert the authenticity of TESLA-enabled GNSS.
    Our algorithms address inherent vulnerabilities with broadcast-only TESLA.
    We modify the standard NTS for network-based time synchronization by removing fields and introducing randomness.
    We introduce checks based on the NTS measurements that are safe to man-in-the-middle delays.
    Using Algorithms~\ref{alg: correct time sync}~and~\ref{alg: mhk receipt}, a receiver can provably be assured that messages were generated by the authenticated provider, subject to the normal cryptographic primitive security assurities.

\section*{Acknowledgments}

\noindent We gratefully acknowledge the support of the FAA Satellite Navigation Team for funding this work under Memorandum of Agreement \#: 693KA8-19-N-00015. Moreover, we gratefully acknowledge the insightful comments by the anonymous reviewers during the peer review of this work.

\bibliographystyle{ieeetran}
\bibliography{references}

\appendix

\subsection{Proof of Potential Forgery with Multicadence TESLA} \label{app: proof}

    \begin{figure}
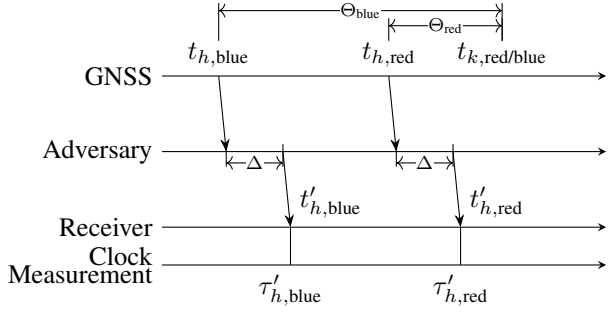

        \def\eplength{0.1}
        \def\eventlength{0.75}
        \tikz{
            \node[anchor=east] (GNSS) at (0, 0) {GNSS};
            \node[anchor=east] (Adv) at (0, -1) {Adversary};
            \node[anchor=east] (Rec) at (0, -2) {Receiver};
            \node[anchor=east, text width=2cm, align=right, font=\linespread{0.5}\selectfont] (RecC) at (0, -2.5) {Clock Measurement};

            \node (GNSSe) at (\lineLength, 0) {};
            \node (Adve) at (\lineLength, -1) {};
            \node (Rece) at (\lineLength, -2) {};
            \node (RecCe) at (\lineLength, -2.5) {};

            \draw[-stealth] (GNSS) -- (GNSSe);
            \draw[-stealth] (Adv) -- (Adve);
            \draw[-stealth] (Rec) -- (Rece);
            \draw[-stealth] (RecC) -- (RecCe);

            \node[connect] (tk) at (6*\eventlength, 0) {};
            \node[connect] (th) at (\eventlength, 0) {};
            \node[connect] (tha) at (\eventlength + \eplength, -1) {};
            \node[connect] (thb) at (2*\eventlength + \eplength, -1) {};
            \node[connect] (thp) at (2*\eventlength + 2*\eplength, -2) {};
            \node[connect] (tauhp) at (2*\eventlength + 2*\eplength, -2.5) {};
            \draw[-{Stealth[scale=1]}] (th) -- (tha);
            \draw[-{Stealth[scale=1]}] (thb) -- (thp);
            \draw[-] (thp) -- (tauhp);

            \node[above] (thl) at (th) {$t_{h, \text{blue}}$};
            \node[above right] at (thp) {$t_{h, \text{blue}}'$};
            \node[below] at (tauhp) {$\tau_{h, \text{blue}}'$};
            \node[above] (tkl) at (tk) {$t_{k, \text{red/blue}}$};

            \draw[-] ([yshift=0.1cm]\eventlength + \eplength, -1) -- ([yshift=-0.2cm]\eventlength + \eplength, -1);
            \draw[-] ([yshift=0.1cm]2*\eventlength + \eplength, -1) -- ([yshift=-0.2cm]2*\eventlength + \eplength, -1);
            \draw[<->] ([yshift=-0.15cm]\eventlength + \eplength, -1) -- node[midway, fill=white, inner sep=0pt] {\scriptsize$\Delta$} ([yshift=-0.15cm]2*\eventlength + \eplength, -1);

            \draw[-] ([yshift=0.4cm]thl.north) -- ([yshift=-0.1cm]thl.north);
            \draw[-] ([yshift=0.4cm]tkl.north) -- ([yshift=-0.1cm]tkl.north);
            \draw[<->] ([yshift=0.3cm]thl.north) -- node[midway, fill=white, inner sep=0pt] {\scriptsize$\Theta_\text{blue}$} ([yshift=0.3cm]tkl.north);

            \node[connect] (th1) at (\eventlength + 3*\eventlength, 0) {};
            \node[connect] (tha1) at (\eventlength + \eplength + 3*\eventlength, -1) {};
            \node[connect] (thb1) at (2*\eventlength + \eplength + 3*\eventlength, -1) {};
            \node[connect] (thp1) at (2*\eventlength + 2*\eplength + 3*\eventlength, -2) {};
            \node[connect] (tauhp1) at (2*\eventlength + 2*\eplength + 3*\eventlength, -2.5) {};
            \draw[-{Stealth[scale=1]}] (th1) -- (tha1);
            \draw[-{Stealth[scale=1]}] (thb1) -- (thp1);
            \draw[-] (thp1) -- (tauhp1);

            \node[above] (thl1) at (th1) {$t_{h, \text{red}}$};
            \node[above right] at (thp1) {$t_{h, \text{red}}'$};
            \node[below] at (tauhp1) {$\tau_{h, \text{red}}'$};

            \draw[-] ([yshift=0.1cm]\eventlength + \eplength + 3*\eventlength, -1) -- ([yshift=-0.2cm]\eventlength + \eplength + 3*\eventlength, -1);
            \draw[-] ([yshift=0.1cm]2*\eventlength + \eplength + 3*\eventlength, -1) -- ([yshift=-0.2cm]2*\eventlength + \eplength + 3*\eventlength, -1);
            \draw[<->] ([yshift=-0.15cm]\eventlength + \eplength + 3*\eventlength, -1) -- node[midway, fill=white, inner sep=0pt] {\scriptsize$\Delta$} ([yshift=-0.15cm]2*\eventlength + \eplength + 3*\eventlength, -1);

            \draw[-] ([yshift=0.2cm]thl1.north) -- ([yshift=-0.1cm]thl1.north);
            \draw[-] ([yshift=0.2cm]tkl.north) -- ([yshift=-0.1cm]tkl.north);
            \draw[<->] ([yshift=0.1cm]thl1.north) -- node[midway, fill=white, inner sep=0pt] {\scriptsize$\Theta_\text{red}$} ([yshift=0.1cm]tkl.north);
        }   
        \caption{
            A conceptual diagram of the attack against a multicadence TESLA receiver that does not meet the synchronization condition of the tighter TESLA instance.
            Latencies are not depicted, but they approach 0.
        }
        \label{fig: multicadence forgery}
    \end{figure}

        \colorlet{darkgreen}{green!50!black}
        \begin{figure}
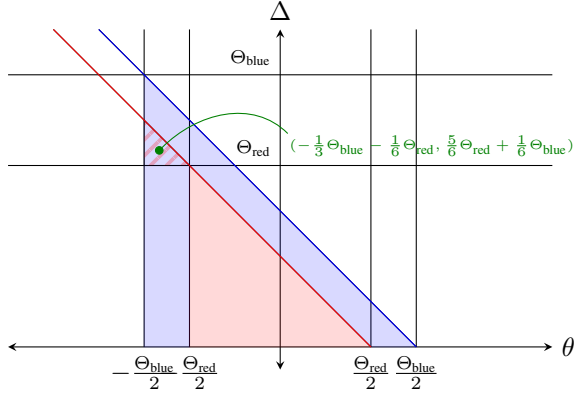

        \centering
        \def\thetalength{0.6}
        \tikz{

            \foreach \i in {-6, -5, -4, -3, -2, -1, 0, 1, 2, 3, 4, 5, 6}{
                \foreach \j in {0, 1, 2, 3, 4, 5, 6, 7}{
                    \node[pt] (p\i\j) at (\i*\thetalength, \j*\thetalength) {};
                }
            }

            \draw[stealth-stealth] (p-60) -- (p60);
            \node[anchor=west] at (p60) {$\theta$};
            \draw[stealth-stealth] ([yshift=-0.3cm]p00.center) -- (p07);
            \node[anchor=south] at (p07) {$\Delta$};

            \node[anchor=north] at (p20) {$\frac{\Theta_\text{red}}{2}$};
            \node[anchor=north] at (p30) {$\frac{\Theta_\text{blue}}{2}$};
            \node[anchor=north] at (p-20) {$-\frac{\Theta_\text{red}}{2}$};
            \node[anchor=north] at (p-30) {$-\frac{\Theta_\text{blue}}{2}$};
            \node[anchor=south east] at (p04) {\scriptsize$\Theta_\text{red}$};
            \node[anchor=south east] at (p06) {\scriptsize$\Theta_\text{blue}$};

            \draw[-] (p20) -- (p27);
            \draw[-] (p30) -- (p37);
            \draw[-] (p-20) -- (p-27);
            \draw[-] (p-30) -- (p-37);
            \draw[-] (p-64) -- (p64);
            \draw[-] (p-66) -- (p66);

            \draw[-] (p-57) -- (p20);
            \draw[-] (p-47) -- (p30);

            \filldraw[blue, opacity=0.15] (p-30.center) -- (p-20.center) -- (p-24.center) -- (p20.center) -- (p30.center) -- (p-36.center) -- cycle;
            \filldraw[red, opacity=0.15] (p-20.center) -- (p20.center) -- (p-24.center) -- cycle;
            \filldraw[pattern={Lines[angle=45, distance={6pt/sqrt(2)}, line width=0.05cm]}, pattern color=red, opacity=0.5] (p-34.center) -- (p-24.center) -- (p-35.center) -- cycle;

            \draw[-, blue] (p-47) -- (p30);
            \draw[-, red] (p-57) -- (p20);

            \node[draw, circle, darkgreen, minimum size=0.1cm, fill, inner sep=0, outer sep=0cm] (centroid) at (-16/6*\thetalength, 26/6*\thetalength) {};
            \node[darkgreen, anchor=south west] (centroidl) at (p04) {\tiny$(-\frac 1 3 \Theta_\text{blue} - \frac 1 6 \Theta_\text{red}, \frac 5 6 \Theta_\text{red} + \frac 1 6 \Theta_\text{blue})$};

            \draw[darkgreen] (centroid) to[in=135, out=45] ([xshift=0.1cm, yshift=-0.1cm]centroidl.north west);

            }
        \caption{
            A diagram showing the unsafe region when a clock certified for one TESLA instance safety region attempts to enforce checks on a tighter TESLA instance.
            The diagram shows regions among the adversary-selected state $\Delta$ and the GIC drift state $\theta$.
            The figure depicts the safety condition lines, and the hashed triangle is the unsafe region where a forgery can occur.
            The centroid of the region is marked as an example state that will induce a forgery.
        }
        \label{fig: insecure triangle}
    \end{figure}

    Figure~\ref{fig: multicadence forgery} provides a conceptual diagram of the attack scenario.
    Suppose we have two TESLA instances, $\Theta_\text{blue}$ and $\Theta_\text{red}$, and $\Theta_\text{red} < \Theta_\text{blue}$.
    To simplify the proof, without loss of generality, we assume that they share the same delayed keys and have a message-commitment-message-commitment-key cadence.
    We examine the scenario where a receiver uses a successful authentication via the blue instance to certify the receipt safety of the red TESLA information.
    To prove no receipt safety, we will show that there exists an adversary induced $\Delta \geq \Theta_\text{red}$ and clock state $\theta$ that is safe for the blue instance \eqref{eq:cond1}, passes the check for the blue instance \eqref{eq:cond4}, passes the check for the red instance \eqref{eq:cond2}, but breaks the receipt safety condition for the red instance \eqref{eq:cond3}.
    In order, this corresponds to
    \begin{align}
        -\frac{\Theta_\text{blue}}{2} &< \theta, \label{eq:cond1} \\
        \tau_{h, \text{blue}}' &< t_k - \frac{\Theta_\text{blue}}{2}, \label{eq:cond4} \\
        \tau_{h, \text{red}}' &< t_k - \frac{\Theta_\text{red}}{2}, \label{eq:cond2} \\
        t_{h, \text{red}}' &\geq t_k. \label{eq:cond3} 
    \end{align}

    From Equation~\eqref{eq:cond2}, we substitute the Equation~\eqref{eq: meas eqn},
    \begin{align}
        \tau_{h, \text{red}}' &< t_k - \frac{\Theta_\text{red}}{2}, \tag{\ref{eq:cond2}} \\
        t_{h, \text{red}}' + \theta &< t_k - \frac{\Theta_\text{red}}{2}, \nonumber \\
        t_{h, \text{red}} + \Delta + \theta &< t_k - \frac{\Theta_\text{red}}{2}, \nonumber \\
        \Delta + \theta &< t_k - t_{h, \text{red}} - \frac{\Theta_\text{red}}{2}, \nonumber \\
        \Delta + \theta &< \frac{\Theta_\text{red}}{2}. \label{eq: cond a}
    \end{align}

    Starting with Equation~\eqref{eq:cond3}, we substitute and simplify,
    \begin{align}
        t_{h, \text{red}}' &\geq t_k, \tag{\ref{eq:cond3}} \\
        t_{h, \text{red}} + \Delta &\geq t_k, \nonumber \\
        \Delta &\geq t_k - t_{h, \text{red}}, \nonumber \\
        \Theta_\text{red} &\leq \Delta . \label{eq: cond b}
    \end{align}

    Combining Equations~\eqref{eq: cond a}~and~\eqref{eq: cond b} yields
    \begin{align}
        \Delta + \theta &< \frac{\Theta_\text{red}}{2}, \tag{\ref{eq: cond a}} \\
        \Delta &< \frac{\Theta_\text{red}}{2} - \theta, \nonumber \\
        \Theta_\text{red} &< \frac{\Theta_\text{red}}{2} - \theta, \nonumber \\
        \theta &< -\frac{\Theta_\text{red}}{2}. \label{eq:final t upper}
    \end{align}

    Combining Equations~\eqref{eq:cond1},~\eqref{eq: cond a},~\eqref{eq: cond b},~and~\eqref{eq:final t upper}, we have the necessary $\Delta$ and $\theta$ to induce a forgery:
    \begin{align}
        \Delta + \theta &< \frac{\Theta_\text{red}}{2}., \tag{\ref{eq: cond a}} \\
        \Theta_\text{red} &\leq \Delta, \tag{\ref{eq: cond b}} \\
        -\frac{\Theta_\text{blue}}{2} &< \theta < -\frac{\Theta_\text{red}}{2}.
    \end{align}
    When generating a plot like Figure~\ref{fig: cert attack}, we observe the triangle of Figure~\ref{fig: insecure triangle}.
    One can verify that each the points in the interior (and some of the boundaries) satisfy Equations~\eqref{eq:cond1}~through~\eqref{eq:cond3}.
    The figure includes the triangle centroid, which will always satisfy the conditions for forgery for the tighter TESLA instance.

\end{document}